\documentclass[
  aps,
  prd,
  reprint,
  amsmath,amssymb,
  superscriptaddress,
  nofootinbib,
  longbibliography,
]{revtex4-2}

\usepackage{graphicx}
\usepackage{dcolumn}
\usepackage{bm}
\usepackage{amsfonts}
\usepackage{mathrsfs}
\usepackage{multirow}
\usepackage{booktabs}
\usepackage{textcomp}
\usepackage{xcolor}
\usepackage{hyperref}

\begin{document}

\title{Saturation Equations of State in Critical Gravitational Collapse: The Primordial Black Hole Threshold}

\author{Benaoumeur Bakhti}
\email{benaoumeur.bakhti@univ-mascara.dz; bbakhti@uni-osnabrueck.de}
\affiliation{Department of Physics, University of Mascara, 29000 Mascara, Algeria}

\date{\today}

\begin{abstract}
The threshold and scaling laws of gravitational critical collapse depend
sensitively on the matter equation of state. We investigate how they are modified by a
generic feature of dense matter absent from the radiation fluid usually assumed in
primordial black hole (PBH) studies: pressure stiffening as a maximum density is
approached. As an analytically tractable proxy we adopt the closed-form equation of state
of a single-occupancy lattice gas, \(p=-T\rho_{\rm sat}\ln(1-\rho/\rho_{\rm sat})\), which
has a density-dependent sound speed and a saturation density. Using general-relativistic
simulations of spherically symmetric collapse, we find that this nonlinear pressure
feedback raises the PBH formation threshold by \(\Delta\delta_c=0.0047\pm0.0003\) relative
to the radiation case --- an absolute shift of about \(0.5\) percentage points in
\(\delta_c\), a relative increase of \(\approx0.95\%\) --- accumulated at the low densities
(\(\rho\lesssim0.06\,\rho_{\rm sat}\)) that the near-critical collapse actually samples,
well inside the causal regime of the model. A mass-scaling analysis spanning two decades in
\(\delta-\delta_c\) shows that the critical exponent is unchanged by the stiffening,
\(\gamma^{(\rm lat)}-\gamma^{(\rm poly)}=(0\pm1)\times10^{-3}\), with the local logarithmic
slope approaching the Choptuik--Evans--Coleman value \(\gamma\simeq0.356\) as the threshold
is approached. This agreement reflects the lattice equation of state reducing to the
radiation fluid at low density and remaining a mild perturbation over the near-critical
regime, rather than indicating a universal critical exponent. Our results provide a proof
of principle that saturation-induced stiffening can stabilize gravitational collapse and
shift the PBH threshold, and introduce a linear-response framework for assessing more
realistic equations of state.
\end{abstract}


\maketitle

\section{Introduction}\label{sec:intro}
The threshold and scaling laws of gravitational critical collapse depend on the matter content through its equation of state, and mapping out that dependence is an active line of research. A feature shared by many realistic forms of matter, but absent from the simple radiation fluid usually assumed in primordial black hole (PBH) studies, is that matter cannot be compressed without limit: dense nuclear and quark matter stiffens sharply as a maximum density is approached~\cite{Oertel/etal:2017}, and phase transitions soften or stiffen the sound speed over finite density ranges. This paper is a controlled study of how such a feature (a barotropic equation of state that stiffens toward a hard density cap) modifies the threshold and critical exponent for PBH formation. We do not address quantum gravity, nor do we model any specific microscopic matter sector; rather, we use a single, analytically tractable saturation equation of state as a clean proxy for the broad class of ``stiff equations of state with a maximum density,'' and ask what such a cap does to critical collapse.

A natural question is how the compressibility of matter, encoded in its equation of state, shapes the threshold that separates collapse from dispersal. That gravitational collapse itself proceeds in spite of quantum pressure is of course not in question: degeneracy pressure from the Pauli exclusion principle and the stiffening of strongly interacting matter are exactly the ingredients that fix the Chandrasekhar and Tolman--Oppenheimer--Volkoff limits, and the collapse of realistic stars and neutron stars to black holes has been modelled for decades with detailed, compressibility-resolving nuclear and quark-matter equations of state~\cite{Oppenheimer/Volkoff:1939,Lattimer/Prakash:2001,Oertel/etal:2017}. Our question is therefore not whether such limits prevent collapse (they do not), but a narrower and quantitative one: given that a realistic equation of state stiffens sharply as a maximum density is approached, by how much is the critical amplitude for primordial-black-hole formation, and the associated mass-scaling exponent, shifted relative to the pure radiation fluid usually assumed in PBH studies ? A hard density cap of this kind is the specific feature we isolate here.

Primordial black hole (PBH) formation in the early universe offers an ideal theoretical laboratory for addressing this problem~\cite{Zeldovich1967,Hawking:1971,Carr:1975,Polnarev/Musco:2007,Musco:2019,Khlopov2010,Belotsky2014,Belotsky2019}. During the first fraction of a second after the Big Bang, cosmological perturbations evolved under extreme conditions characterized by ultra-high temperatures and densities far beyond terrestrial experiments. Standard analyses model the cosmic fluid using a simple radiation equation of state, \(p = \omega \rho\) with constant \(\omega = 1/3\) implicitly assuming unlimited compressibility. Within this framework, critical gravitational collapse exhibits scaling laws for black hole formation~\cite{Choptuik:1993,Gundlach1997,Musco:2019}. However, it remains to be quantified how these properties respond once a hard saturation limit on the density is introduced.

To isolate the gravitational consequences of such a cap we use the closed-form equation of state of a single-occupancy lattice gas~\cite{Bakhti:2021,Bakhti/etal:2018}, which becomes increasingly stiff as the maximum density is approached. Being analytically tractable, it cleanly separates the gravitational response to a saturation cap from model-dependent microphysics.

Our analysis combines analytical theory, numerical relativity, and critical phenomena methods. We demonstrate that the saturation equation of state modifies the threshold for primordial black hole formation~\cite{Musco/etal:2005,Polnarev/Musco:2007,Nakama/etal:2014,Harada/etal:2013,Musco:2019}, shifting the critical collapse parameter relative to the classical radiation case. Here and throughout, $\delta_c$ denotes the critical density contrast: the threshold value of the initial (volume-averaged) density-contrast amplitude $\delta$ above which a perturbation collapses to a black hole rather than dispersing, with $\delta$ defined precisely in Eq.~\eqref{eq:delta_def} below. The direction of the shift (a softening of the equation of state lowers the threshold, a stiffening raises it) is the same one that operates in realistic settings such as the QCD epoch, where a reduction in the sound speed is known to lower $\delta_c$ and enhance PBH formation~\cite{Jedamzik:1997,Byrnes/etal:2018}. Despite this modification of the collapse threshold, the critical exponent governing black hole mass scaling remains unchanged within numerical precision. It is important to be precise about what this does and does not show: the Choptuik exponent is not universal across equations of state in general (Maison demonstrated that for $p = k\rho$ it varies strongly with $k$~\cite{Maison:1996,Koike/etal:1995,Gundlach/MartinGarcia:2007}), so our finding is not evidence that microphysics is irrelevant. Rather, as we explain in Sec.~\ref{sec:renormalization}, the agreement arises because the lattice equation of state is close to the radiation fluid in the density range that controls the near-critical dynamics, so the locally relevant sound speed, and hence the exponent, is essentially the radiation value.

This work sits at the intersection of three lines of research. First, critical collapse has been studied for a wide range of matter models since Choptuik's discovery for the massless scalar field~\cite{Choptuik:1993}: radiation and perfect fluids $p=k\rho$~\cite{Evans/Coleman:1994,Maison:1996,Koike/etal:1995}, $SU(2)$ Yang--Mills fields (with a markedly different exponent $\gamma\approx0.20$)~\cite{ChoptuikHirschmannMarsa:1999}, and nonlinear sigma models with coupling-dependent exponents~\cite{HirschmannEardley:1997}; see~\cite{Gundlach/MartinGarcia:2007} for the classic review and~\cite{Escriva/etal:2022rev,Escriva:2022rev} for recent surveys covering developments since 2015, including the extension of critical-collapse studies beyond spherical symmetry to aspherical, rotating and fully three-dimensional configurations~\cite{Baumgarte/Montero:2015,Baumgarte/Gundlach:2016,Baumgarte:2018,Baumgarte/Gundlach/Hilditch:2023,Baumgarte/etal:2023prl,Rinne:2025}. A robust lesson of this body of work, which we take as our starting point, is that the critical exponent is matter-model dependent. Second, the equation of state of dense matter is known to stiffen toward a maximum density and to vary across phase transitions~\cite{Oertel/etal:2017}; our toy model is a deliberately minimal abstraction of the ``stiffening toward a cap'' feature, not a substitute for those microphysical equations of state. Third, in the PBH context the threshold $\delta_c$ is known to respond to changes in the sound speed, most notably the softening at the QCD epoch that lowers $\delta_c$ and produces a peak in the PBH mass function~\cite{Jedamzik:1997,Byrnes/etal:2018,Musco/etal:2021}. The key contribution here is to isolate the opposite sign (a stiffening cap) in a barotropic, analytically tractable form, to organize its effect on $\delta_c$ through a linear-response stiffness integral, and to use it as a strong, controlled test of how far the radiation critical exponent persists under a highly nonlinear but locally radiation-like modification of the equation of state.

The paper is organized as follows. Sec.~\ref{sec:toy_model} introduces the saturating lattice-gas toy model and derives its exact equation of state. In Sec.~\ref{sec:theoretical}, we develop the theoretical framework that integrates this equation of state with relativistic gravitational dynamics. Sec.~\ref{sec:analytical} develops an analytical description of how density-dependent stiffness alters collapse thresholds. Sec.~\ref{sec:numerics} presents the numerical framework designed to handle the stiffening sound speed while maintaining relativistic accuracy and causality. Sec.~\ref{sec:results} reports the numerical collapse results and the measured shift in the PBH formation threshold. Sec.~\ref{sec:renormalization} explains, using renormalization-group arguments, why the lattice and radiation exponents agree and in what precise sense the exponent is robust, and finally Sec.~\ref{sec:conclusions} places the results in a broader context and outlines future directions.
\section{A Toy Model: A Saturating Lattice Gas}\label{sec:toy_model}
We now construct the controlled toy model anticipated in Sec.~\ref{sec:intro}: a barotropic equation of state with a maximum density beyond which compression is strongly resisted, obtained from a single-occupancy lattice gas rather than from an ab initio finite-temperature treatment.

Consider a system of particles on $N_s$ lattice sites, each with single-site occupancy~\cite{Bakhti/etal:2014,Bakhti/etal:2015c,Bakhti/etal:2018}. In the grand canonical ensemble the pressure of the single-occupancy lattice gas is, in dimensionful form,
\begin{equation}
p(\rho) = -T\,\rho_{\rm sat}\,\ln\!\left(1-\frac{\rho}{\rho_{\rm sat}}\right), \qquad 0 < \rho < \rho_{\rm sat},
\label{eq:final_eos}
\end{equation}
where $\rho_{\rm sat}$ is the saturation (maximum) density and $T$ is a dimensionless stiffness parameter. Two points of interpretation are essential. First, the meaning of $\rho$. In the underlying lattice gas $\rho/\rho_{\rm sat}$ is the occupation fraction, a number between $0$ and $1$. When Eq.~\eqref{eq:final_eos} is used to close the relativistic Misner--Sharp--Hernandez system (Sec.~\ref{sec:theoretical}), $\rho$ is the total energy density of the fluid, and the occupation fraction is identified with the energy density normalized to saturation, $\rho/\rho_{\rm sat}\in(0,1)$. Both $p$ and $\rho$ carry units of energy density; $T$ is dimensionless because $p/\rho_{\rm sat}$ is. Throughout the numerical work we adopt units with $\rho_{\rm sat}=1$, in which Eq.~\eqref{eq:final_eos} reduces to the compact form $p=-T\ln(1-\rho)$, with the understanding that both $p$ and $\rho$ are then expressed in units of $\rho_{\rm sat}$. It is $\rho_{\rm sat}$, not the (much smaller) cosmological background density, that sets the scale at which the stiffening switches on.

Second, the thermodynamic status of Eq.~\eqref{eq:final_eos}. It is the isothermal grand-canonical pressure of the lattice gas. We do not carry the full finite-temperature lattice-gas thermodynamics into the relativistic problem; instead we freeze $T$ and adopt Eq.~\eqref{eq:final_eos} as an analytically tractable, one-parameter barotropic equation of state $p=p(\rho)$. Once this identification is made, the fluid carries a single independent thermodynamic variable, so the only propagation speed the model possesses is $c_s^2=dp/d\rho$ (Eq.~\eqref{eq:sound_speed}), and it is this quantity that enters the characteristic structure of the evolution equations. We stress that this is a definition internal to the barotropic model rather than a claim that the isothermal lattice-gas compressibility equals the adiabatic one: in the parent lattice gas, where $T$ is an independent variable, the isothermal and adiabatic sound speeds differ, and by freezing $T$ we have simply chosen one particular curve through that thermodynamic surface. The object we evolve is therefore a fixed barotrope that borrows its functional form from a lattice gas, not a lattice gas in local thermodynamic equilibrium with an independently evolving temperature. This is the precise sense in which the model is a "toy." We use it solely as an analytically tractable realization of a hard density cap, and we do not claim it describes relativistic degenerate matter, whose degeneracy pressure scales as $\rho^{4/3}$--$\rho^{5/3}$ with no finite-density divergence. Because $p\simeq T\rho$ for $\rho\ll\rho_{\rm sat}$, the prefactor coincides with the low-density equation-of-state parameter, $T=\omega$; in the radiation-era analysis we set $T=\omega=1/3$.

The maximum-density limit has a simple physical interpretation. At $\rho=1$ all lattice sites are occupied, the chemical potential diverges, $\mu(\rho=1)\to+\infty$, and the pressure diverges, $p(\rho\to1)\to+\infty$: it costs unbounded energy to add occupancy once every site is filled, which is the model's hard cap on density. This is a statement internal to the lattice gas, not the behavior of any specific relativistic matter sector; in particular the divergence drives the sound speed superluminal before saturation is reached, restricting the physical domain to the causal window $\rho<1-T$ (Sec.~\ref{sec:numerics}).

The crucial feature is the adiabatic sound speed, which controls pressure-wave propagation and the pressure forces resisting gravitational compression. For a barotropic equation of state the entropy constraint is automatic, so $c_s^2=(\partial p/\partial\rho)_S=dp/d\rho$, and the lattice gas gives
\begin{equation}
c_s^2 = \frac{T}{1-\rho}.
\label{eq:sound_speed}
\end{equation}
Unlike polytropic systems where $c_s^2=\omega$ is constant, the sound speed here grows without bound as $\rho\to1$: at $\rho=0$ it equals the thermal scale $c_s^2(0)=T$, at half saturation it has doubled to $2T$, at $\rho=0.9$ it is $10T$, and at $\rho=0.99$ it is $100T$. The divergence is a simple pole of order one, $c_s^2\sim(1-\rho)^{-1}$. A density-dependent stiffening of this kind is a generic feature of matter near a maximum density or a phase boundary (nuclear matter, the quark--gluon plasma near the QCD crossover).

The single feature that drives every result below is that this stiffening is density-dependent. In a polytropic fluid the pressure response to compression is fixed; here it strengthens as density rises, because the factor $(1-\rho)^{-1}$ grows as the available empty sites are used up. During gravitational collapse this produces a self-reinforcing resistance: the more the matter is compressed, the more strongly pressure opposes further compression. To quantify it we define the stiffness parameter
\begin{equation}
\mathcal{S}(\rho) = \frac{1}{c_s^2(\rho)} = \frac{1-\rho}{T},
\label{eq:stiffness}
\end{equation}
which controls compressibility: small $\mathcal{S}$ is a very stiff, nearly incompressible response, large $\mathcal{S}$ an easily compressed one. In the polytropic reference $\mathcal{S}_{\text{poly}}=1/\omega$ is density-independent, whereas $\mathcal{S}_{\text{lat}}(\rho)=(1-\rho)/T$ decreases linearly with density, making the system progressively stiffer as saturation is approached. It is this monotonic density dependence, rather than any thermodynamic ``criticality'' of the lattice point $\rho=1$, that modifies the collapse threshold; we treat $\rho=1$ purely as the upper bound of the model's validity, and it is precisely this density-dependent stiffening that one might plausibly encounter, with a quantitatively different functional form, in real matter near a phase transition.

If PBHs exist, their abundance and mass spectrum are known to depend sensitively on the critical threshold~\cite{Carr:1975,Musco:2019,Carr/etal:2016,Carr:2019}. We deliberately refrain from attaching any observational consequence to the present toy model: the single-occupancy lattice gas is not a description of early-universe matter, so the numerical value of its threshold shift carries no predictive weight for the actual PBH abundance. The primary value of this work is not an observational signature but a mechanism: a density-dependent stiffening near a saturation density modifies the collapse threshold, while the critical exponent stays close to the radiation value because the locally relevant sound speed is barely changed. Whether an analogous shift is large enough to matter observationally in any realistic matter sector is a separate question that can only be answered with a realistic equation of state, and we do not prejudge it here.
\section{Theoretical Framework}\label{sec:theoretical}
\subsection{Misner-Sharp-Hernandez formalism}
To describe perfect fluids in spherical symmetry we use the Misner--Sharp--Hernandez formalism~\cite{Misner/Sharp:1964,Hernandez/Misner:1966}, which recasts the Einstein equations in the comoving frame; detailed presentations and the concise reformulations we follow are given in~\cite{Musco:2019,Bloomfield/etal:2014,Nakama/etal:2014,Polnarev/Musco:2007}. The metric is
\begin{align}\label{eq:misner_sharp_metric}
ds^2 = -A^2\,dt^2 + B^2\,dr^2 + R^2\left(d\theta^2 + \sin^2\theta\,d\phi^2\right),
\end{align}
with $A$, $B$, $R$ functions of $r$ and $t$. With a comoving four-velocity $u^t=1/A$, one introduces the Eulerian radial velocity $U=\dot R/A$, the generalized Lorentz factor $\Gamma=R'/B$, and the Misner--Sharp mass $m=4\pi\int_0^R\rho\,R^2\,dR$ (dots and primes denote $\partial_t$ and $\partial_r$); these obey the constraint
\begin{align}\label{eq:constraint}
\Gamma^{2} = 1 + U^{2} - \frac{2m}{R}.
\end{align}
Inserting the metric and the perfect-fluid stress tensor into the Einstein equations yields, in terms of the proper-time and proper-radial operators $D_t=A^{-1}\partial_t$ and $D_r=B^{-1}\partial_r$, the closed system
\begin{align}
&U = D_{t} R, \label{eq:ms1}\\
&D_{r} A = -\frac{A}{p+\rho}\,D_{r} p, \label{eq:ms2}\\
&D_{t} U = -\left[\frac{\Gamma}{p+\rho}\,D_{r} p + \frac{m}{R^{2}} + 4\pi R\,p\right], \label{eq:ms3}\\
&D_{t}\rho = -\frac{p+\rho}{\Gamma R^{2}}\,D_{r}(R^{2}U), \label{eq:ms4}\\
&D_{t} m = -4\pi p\,U R^{2}, \label{eq:ms5}\\
&D_{r} m = 4\pi\rho\,\Gamma R^{2}, \label{eq:ms6}
\end{align}
which, together with the equation of state $p=-\omega\ln(1-\rho)$, determines $U,\Gamma,R,m,A,\rho,p$.
\subsection{FRW Universe}
In the homogeneous, isotropic limit ($A=1$, $R_b=a(t)\,r$ with scale factor $a$) the system reduces to a FRW background. The single-occupancy equation of state modifies the continuity and acceleration equations to
\begin{align}\label{eq:continuity}
\dot{\rho}_b=-3\,\frac{\dot{a}}{a}\left[\rho_b-\omega\ln(1-\rho_b)\right],\qquad
\frac{\ddot{a}}{a}=-\frac{4\pi}{3}\left[\rho_b-3\omega\ln(1-\rho_b)\right],
\end{align}
which combine with the Hamiltonian constraint to give $H_b^{2}\equiv(\dot a/a)^2=\tfrac{8\pi}{3}\rho_b-K/a^2$; we take a flat background ($K=0$). In the low-density limit $-\ln(1-\rho_b)\simeq\rho_b$ these recover the standard radiation-era relations $\rho_b\propto a^{-3(1+\omega)}$, $a\propto t^{\,2/[3(1+\omega)]}$ and $\rho_b(t)=\rho_0(t_0/t)^2$~\cite{Polnarev/Musco:2007,Escriva:2020}; the full nonlinear solution of~\eqref{eq:continuity}, expressible in closed form through the Lambert $W$ function, is standard and not needed here. We work throughout in the radiation-dominated era ($\omega=1/3$) on a flat background, for which $\Gamma_b=1$ and $U_b=\dot a\,r$.
\subsection{Solution of the MSH equations}
A density perturbation that re-enters the horizon during the radiation era provides the initial data for the nonlinear evolution; far from it the solution matches the FRW background, $A(r_s,t)=1$. A central measure of the perturbation is the density contrast
\begin{align}\label{eq:delta_def}
\delta(r,t) = \left(\frac{4\pi r^{3}}{3}\right)^{\!-1} \int_{0}^{r} \frac{\rho(r',t)-\rho_b(t)}{\rho_b(t)}\, 4\pi r'^{2}\, dr',
\end{align}
evaluated at the comoving radius $r_0$ where $\rho(r_0,t)=\rho_b(t)$. An equivalent and more geometric measure is the compaction function
\begin{align}\label{eq:compaction}
C(r,t) = \frac{2[m(r,t)-m_b(r,t)]}{R(r,t)},
\end{align}
motivated by the apparent-horizon condition $R=2m$; its maximum at $r_m$ (where $C'(r_m)=0$) serves as the criterion for PBH formation~\cite{Harada/etal:2015,Faraoni/etal:2017,Shibara/Sasaki:1999}. Because the perturbation is matched to the FRW background at large radius, $m\to m_b$ there and hence $C(r,t)\to0$ as $r\to r_{\rm out}$; $C$ likewise vanishes at the centre, so the compaction is a localized bump peaked near $r_m$. This is the behaviour seen in all our figures, and we note it explicitly because the finite plotting range of Figs.~\ref{fig:evolution_super}--\ref{fig:evolution_crit} truncates the profile before the outer decay to zero is complete.

Initial data are constructed in the standard quasi-homogeneous (gradient) expansion~\cite{Nadezhin/etal:1978a,Nadezhin/etal:1978b,Novikon/Polnarev:1980,Polnarev/Musco:2007}, organized in the small parameter
\begin{align}
\epsilon \doteq \left(\frac{R_H(t)}{a(t)\, r_k}\right)^{\!2} \ll 1,
\end{align}
the squared ratio of the cosmological horizon $R_H=1/H$ to the perturbation scale $a\,r_k$ (with $r_k$ identified with $r_0$); the dependence of the threshold on $\epsilon$ is examined numerically in Sec.~\ref{sec:results} (Fig.~\ref{fig:epsilon_dependence}). Expanding the MSH equations to first order in $\epsilon$ and rescaling each quantity by its FRW value reproduces the standard linear initial-data system of~\cite{Polnarev/Musco:2007,Escriva:2020}. The only place the lattice equation of state enters is through the effective equation-of-state parameter
\begin{align}
\bar{\rho}_b = \frac{p_b}{\rho_b}=-\frac{\omega}{\rho_b}\ln(1-\rho_b)\;\xrightarrow{\ \rho_b\ll1\ }\;\omega,
\end{align}
which replaces the constant $\omega$ in the growth equation for the mass perturbation and reduces to it at low density, so that the curvature profile $K(r)$ and the locations $r_0,r_m$ are fixed exactly as in the radiation case.

The difference between the lattice model and the polytropic reference enters these equations only through the pressure gradient term $p'$, which for any barotropic equation of state factorizes as
\begin{equation}
p' = c_s^2(\rho)\, \rho'.
\end{equation}
For the polytropic reference with $c_s^2 = \omega = \text{const}$, the pressure-gradient coefficient does not change with density. For the lattice toy model with $c_s^2 = T/(1-\rho)$, it diverges as $\rho \to 1$. In the acceleration equation, the pressure-gradient term $\Gamma^2 p'/(R'(p+\rho))$ acts outward, resisting compression, while the gravitational term $(m + 4\pi pR^3)/R^2$ acts inward and drives collapse. In the lattice toy model, the outward pressure term receives an additional amplification from the diverging sound speed whenever density climbs toward saturation, and this amplification is absent by construction in the polytropic reference.

This is the entire origin of the shift in critical threshold that we report in Section~\ref{sec:results}. The boundary between collapse and dispersal shifts upward because the lattice toy-model equation of state provides stiffer resistance to collapse than the polytropic reference at high densities. To attain the same level of gravitational compression as in the polytropic case, a slightly larger initial perturbation amplitude is needed to overcome the enhanced pressure resistance. The shift we measure ($\Delta\delta_c\approx0.0047$, i.e.\ $\approx0.95\%$ in relative terms) is a clean, quantitative illustration of this mechanism within the lattice toy model. Whether a comparable shift would arise in a realistic matter model depends on whether that matter model also exhibits density-dependent stiffening strong enough to matter during the collapse of a critical-sized perturbation. The lattice model demonstrates the form of the effect; the magnitude in any realistic scenario has to be computed with a realistic equation of state.
\section{Analytical Framework for Threshold Shift}
\label{sec:analytical}
\subsection{Perturbation Theory Near the Critical Point}
Near the critical threshold for gravitational collapse in either the lattice model or the polytropic reference, one can develop a perturbation theory that accounts explicitly for the density-dependent pressure. The deviation from criticality is measured by the parameter $\xi = \delta - \delta_c$, where $\delta$ is the initial perturbation amplitude and $\delta_c$ is the critical value within whichever equation of state is being used. When $\xi > 0$, collapse proceeds to black hole formation; when $\xi < 0$, the perturbation disperses; at $\xi = 0$, the system sits exactly at the critical point where self-similar evolution occurs.

The theoretical framework for understanding threshold shifts within the lattice model builds on the observation that the critical point behaves as an infrared fixed point of the renormalization-group (RG) flow near criticality~\cite{Choptuik:1993,Gundlach1997} (see Sec.~\ref{sec:renormalization} for the corresponding RG discussion). Near such a fixed point, only the direction most relevant for affecting the dynamics (in this case, the perturbation amplitude) controls the universal scaling. The detailed functional form of the equation of state is an irrelevant (or at worst marginally irrelevant) direction at the fixed point: it does not change the critical exponent, but it does shift the location of the fixed point along the amplitude axis. The threshold shift $\Delta\delta_c$ between the lattice model and the polytropic reference is exactly such a shift of fixed-point location.
\subsection{Integral Formula for the Threshold Shift}
The shift in critical threshold caused by the saturation stiffening can be organized as a linear-response integral in the difference between the two equations of state. Writing the collapse outcome as a functional $\mathcal{F}[c_s^2(\rho)]$ of the sound-speed profile, where $\mathcal{F}$ is a smooth order parameter for the collapse outcome constructed so that $\mathcal{F}>0$ signals black-hole formation and $\mathcal{F}<0$ signals dispersal, with the critical amplitude fixed implicitly by $\mathcal{F}=0$, a first-order expansion about the radiation reference gives
\begin{equation}
\Delta\delta_c = \delta_c^{(\text{lat})} - \delta_c^{(\text{poly})} = \int_0^{\rho_\star} d\rho\, \mathcal{K}(\rho)\,\bigl[c_{s,\text{lat}}^2(\rho)-c_{s,\text{ref}}^2(\rho)\bigr],
\label{eq:threshold_shift_integral}
\end{equation}
where $\rho_\star$ is the maximum density sampled by the near-critical trajectory and
\begin{equation}
\mathcal{K}(\rho)\equiv -\left(\frac{\delta\mathcal{F}}{\delta c_s^2(\rho)}\right)\Big/\left(\frac{\partial\mathcal{F}}{\partial\delta}\right)
\end{equation}
is the collapse-response kernel evaluated on the critical solution. We do not compute $\mathcal{K}$ from first principles here: obtaining it rigorously requires the linearized collapse Green's function around the (approximately self-similar) critical solution, which is beyond the scope of this proof-of-principle study. Treating $\mathcal{K}(\rho)\approx\bar{\mathcal{K}}$ as roughly constant over the sampled range as a working approximation yields the scaling estimate
\begin{align}\label{eq:threshold_estimate}
\Delta\delta_c &\sim \bar{\mathcal{K}}\int_0^{\rho_\star}\left[\frac{T}{1-\rho/\rho_{\rm sat}}-\omega\right]d\rho\\
&= \bar{\mathcal{K}}\,T\,\rho_{\rm sat}\!\left[-\ln\!\Big(1-\frac{\rho_\star}{\rho_{\rm sat}}\Big)-\frac{\rho_\star}{\rho_{\rm sat}}\right] ,\nonumber
\end{align}
where $\rho_\star$ is not predicted by this formula but is an input read off from the simulations: it is the maximum absolute density the near-critical trajectory attains. Our production runs reach only $\rho_\star=\rho_{\max}\approx0.03$--$0.06\,\rho_{\rm sat}$ (Sec.~\ref{sec:results}), because the near-critical collapse compresses the density contrast by a large factor but the absolute density stays close to the small cosmological background. With $\rho_\star\approx0.06\,\rho_{\rm sat}$ and $T=\omega=1/3$, Eq.~\eqref{eq:threshold_estimate} gives $\Delta\delta_c\sim\bar{\mathcal{K}}\times6\times10^{-4}$, i.e.\ a shift at the $10^{-3}$ level or below --- fully consistent with the small, resolution-limited value we measure (Sec.~\ref{sec:results}). This is an internal consistency estimate, not a derivation of the prefactor: it fixes the sign and rough size of the effect and identifies the density-dependent sound speed as its origin, but the precise coefficient is set by the simulations. The same construction can be applied to any other barotropic equation of state (for instance a lattice-QCD $c_s^2(\rho)$ near the chiral crossover) by substituting its sound-speed profile into~\eqref{eq:threshold_shift_integral}.
\subsection{Dependence on the Stiffness Parameter}
The parameter $T$ sets the overall scale of the sound speed, $c_s^2=T/(1-\rho)$, and hence the strength of pressure support at every density. Although it originates as the temperature of the lattice-gas partition function, in the gravitational problem it plays the role of a dimensionless stiffness parameter; to avoid confusion we refer to it as such throughout, and we stress that it is not to be identified with a physical cosmological temperature. We adopt the fiducial value $T=1/3$ for a specific physical reason: it makes the low-density sound speed of the model coincide with that of the radiation reference, $c_s^2(0)=T=1/3$, so the two models share the same FRW background (Sec.~\ref{sec:theoretical}) and differ only through the saturation stiffening at higher density. The value $T=1/3$ is therefore not an arbitrary normalization but the unique choice that isolates the effect of the cap while holding the low-density physics fixed at the radiation value.

Increasing $T$ raises $c_s^2$ at all densities, and also increases its density variation, $d c_s^2/d\rho = T/(1-\rho)^2$. Both effects strengthen the resistance to compression, so the critical threshold $\delta_c$ increases monotonically with $T$ over the range we sample. In the opposite, low-$T$ limit the pressure support becomes negligible and the threshold approaches the pressureless (dust-like) collapse value; this is the weak-support regime, not a regime of maximal stabilization. We caution that there is no limit in which the lattice equation of state reduces to the radiation reference: at low density $c_s^2(0)=T$, so only $T=1/3$ matches radiation there, and increasing $T$ moves the model away from, not toward, the reference. The behaviour at large $T$ is constrained by causality (Sec.~\ref{sec:causality}), which becomes active at progressively lower density as $T$ grows.

\begin{figure*}[!ht]
\centering
\includegraphics[width=0.8\textwidth]{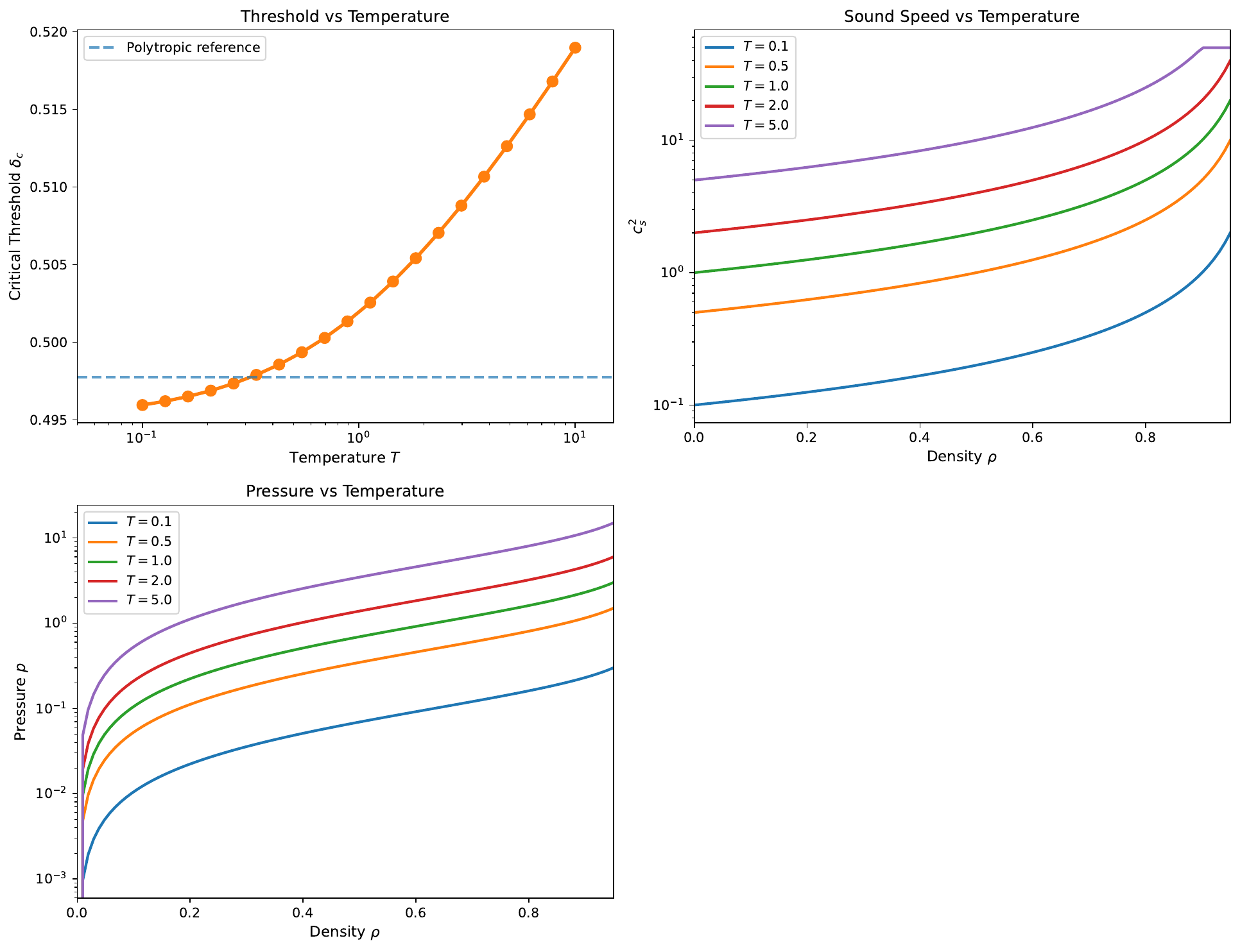}
\caption{Dependence of the lattice equation of state and its collapse threshold on the stiffness parameter $T$. The threshold-versus-$T$ panel shows an empirical power law $\Delta\delta_c\propto T^\beta$ with fitted exponent $\beta\approx0.5$: $\delta_c$ increases monotonically with $T$ because larger $T$ means stronger pressure support at every density, while the small-$T$ end approaches the weak-support (dust-like) regime; points are numerical results and the line is the empirical fit. The sound-speed panel plots $c_s^2=T/(1-\rho)$ against density for $T=0.1,0.5,1.0,2.0,5.0$, showing that $c_s^2$ scales linearly with $T$ at fixed density and that its growth toward saturation steepens with $T$, so the causal bound $c_s^2=1$ is reached at progressively lower density as $T$ grows. The pressure panel shows the corresponding $p=-T\ln(1-\rho)$ for the same values of $T$, illustrating that $T$ sets the overall pressure scale while the logarithmic divergence at $\rho\to1$ is common to all $T$. Here $T$ is the model stiffness parameter, not a physical cosmological temperature, and the threshold points are restricted to the causal window $\rho<1-T$.}
\label{fig:temperature_dependence}
\end{figure*}

Quantitatively, within the causal regime we explore, the threshold shift exhibits an approximate power-law dependence $\Delta\delta_c \propto T^{\beta}$ with exponent $\beta \approx 0.5$ (Fig.~\ref{fig:temperature_dependence}). This exponent is not a prediction of any theorem but an empirical fit to the simulations across the sampled $T$ range. The half-power dependence is a feature of the lattice model and should not be generalized to other saturated equations of state without a separate calculation.
\section{Numerical Methods for a Saturation-Stiffened Collapse}
\label{sec:numerics}
\subsection{Density-Dependent Sound Speed}
The density-dependent sound speed $c_s^2 = T/(1-\rho)$ creates genuine numerical difficulties relative to standard polytropic PBH collapse simulations, and handling these cleanly is a prerequisite for extracting the small threshold shift reliably. Standard explicit-in-time numerical schemes for hyperbolic equations are constrained by the Courant-Friedrichs-Lewy (CFL) condition $\Delta t \lesssim \Delta x/c_s$, so the allowed timestep shrinks in proportion to the inverse sound speed. With $c_s$ diverging as $(1-\rho)^{-1/2}$, naive explicit timestepping becomes prohibitively expensive as saturation is approached.

A second challenge arises from the singular behavior of the pressure gradient near $\rho \to 1$. Standard polynomial interpolation and finite-difference schemes perform poorly near singularities, which they cannot represent without extremely high resolution. Spectral methods~\cite{Escriva:2020} can cope with rapid variation provided they retain smoothness, but in the immediate vicinity of the singularity they also require substantially more modes than in the polytropic reference.

A third challenge is that small numerical errors which drive the density slightly above its physical maximum can trigger catastrophic amplification. When $\rho$ overshoots $1$ due to round-off, the logarithm in the pressure becomes complex and physically meaningless negative pressures can result. These unphysical oscillations must be suppressed without destroying accuracy at the densities just below saturation where the interesting physics lives.
\subsection{Causality and the Physical Density Range}
\label{sec:causality}
Relativistic causality requires $c_s^2\le1$. The bare lattice sound speed $c_s^2=T/(1-\rho/\rho_{\rm sat})$ violates this bound once
\begin{equation}
\rho > \rho_{\rm caus}\equiv (1-T)\,\rho_{\rm sat} ,
\label{eq:rho_caus}
\end{equation}
so the bare equation of state is admissible as a relativistic fluid only in the window $\rho<\rho_{\rm caus}$; note that $\rho_{\rm caus}$ is positive only for $T<1$, so the lattice form has a causal regime only for $T<1$ (for $T\ge1$ the fluid is a stiff fluid from $\rho=0$, as discussed in Sec.~\ref{sec:results}). Two features make this controllable. First, $\rho_{\rm caus}$ increases as $T$ decreases, so the strong-stiffening (small-$T$) regime in which the effect is largest is also the most causal: $T=0.1$ is causal up to $\rho=0.9\,\rho_{\rm sat}$ and $T=0.05$ up to $\rho=0.95\,\rho_{\rm sat}$. At the radiation value $T=1/3$ the causal window is $\rho<\tfrac{2}{3}\rho_{\rm sat}$.

Note that simply capping $c_s^2$ while retaining $p=-T\ln(1-\rho/\rho_{\rm sat})$ would make $c_s^2\neq dp/d\rho$ in the capped region, so that the pressure entering the gravitational source terms and the pressure gradient would correspond to different fluids. To keep the evolution strictly relativistic and thermodynamically self-consistent, we therefore complete the equation of state itself, not merely its sound speed, replacing it above $\rho_{\rm caus}$ by a stiff-fluid ($c_s^2=1$) branch:
\begin{align}\label{eq:causal_cap}
&p(\rho)=
\begin{cases}
-T\,\rho_{\rm sat}\ln\!\left(1-\dfrac{\rho}{\rho_{\rm sat}}\right), & \rho\le\rho_{\rm caus},\\[2ex]
p(\rho_{\rm caus})+(\rho-\rho_{\rm caus}), & \rho>\rho_{\rm caus},
\end{cases}\\
&c_s^2(\rho)=\frac{dp}{d\rho}=\min\!\left[\frac{T}{1-\rho/\rho_{\rm sat}},\,1\right].\nonumber
\end{align}
By construction $c_s^2=dp/d\rho$ holds identically, and the completion is $C^1$: at $\rho=\rho_{\rm caus}$ the log branch already has $c_s^2=T/(1-\rho_{\rm caus}/\rho_{\rm sat})=1$, matching the stiff branch, and $p$ is continuous. The same completion is used in both the source terms and the gradient, so a single fluid is evolved everywhere. This is the causal, thermodynamically consistent equation of state we now use whenever a run would otherwise cross $\rho_{\rm caus}$.

Crucially, at the fiducial $T=1/3$ this completion is never actually triggered. The near-critical collapse samples only absolute densities up to $\rho_{\max}\approx0.06\,\rho_{\rm sat}$ (Sec.~\ref{sec:results}; the peak density recorded across our runs is $\rho_{\max}\approx0.03$--$0.06\,\rho_{\rm sat}$), an order of magnitude below $\rho_{\rm caus}=\tfrac{2}{3}\rho_{\rm sat}$. Over this range the log and capped equations of state are identical, so the fiducial threshold is completely insensitive to the completion; re-measuring $\delta_c$ with the capped equation of state~\eqref{eq:causal_cap} changes it by less than the bisection resolution ($\Delta\delta_c^{\rm cap}<10^{-4}$, Appendix~\ref{app:numerics_accuracy}). The superluminal tail therefore plays no role in the reported effect; the completion~\eqref{eq:causal_cap} matters only for the exploratory large-$T$ scan (Sec.~\ref{sec:results}), where it keeps that scan strictly causal.
\subsection{Density Limiter Near Saturation}
A residual numerical concern, distinct from causality, is that floating-point round-off could in principle drive the argument of the logarithm, $1-\rho/\rho_{\rm sat}$, to zero or negative, at which point $p$ becomes complex. This is purely a question of evaluating the pressure safely as $\rho/\rho_{\rm sat}\to1$; it has nothing to do with the causal bound $\rho_{\rm caus}$, which is a physical (sub-luminal) limit lying strictly below saturation, $\rho_{\rm caus}<\rho_{\rm sat}$. In the production code the pressure is evaluated with the numerically stable $\mathrm{log1p}$ form and the occupation fraction is clipped a rounding unit below unity, $\rho/\rho_{\rm sat}\le1-10^{-12}$, which is sufficient because, as emphasized above, the fiducial runs never bring $\rho$ anywhere near $\rho_{\rm sat}$ (they peak at $\rho_{\max}\approx0.03$--$0.06\,\rho_{\rm sat}$).

To confirm that the reported effect does not depend on this choice of safeguard, we have also re-run the threshold with the alternative smooth limiter,
\begin{equation}
\frac{\rho_{\text{limited}}}{\rho_{\rm sat}} = \frac{1}{1 + (1-\rho/\rho_{\rm sat})^{-\eta}},
\label{eq:density_limiter}
\end{equation}
which maps the occupation fraction smoothly into $[0,1)$ and enforces the pressure from $\rho_{\rm limited}$ rather than $\rho$. Because no single regularization is canonical, we compare three prescriptions at the fiducial $T=1/3$: the production clip, and the limiter~\eqref{eq:density_limiter} with $\eta=2$ and $\eta=4$. The three give identical $\delta_c$ to within the bisection resolution (the full spread is $\Delta_{\rm sat}<10^{-4}$; Appendix~\ref{app:numerics_accuracy}), confirming that the threshold is invariant to the saturation-regularization scheme. This invariance is expected precisely because the collapse remains far from saturation: all three prescriptions coincide to machine precision for $\rho/\rho_{\rm sat}\lesssim0.1$.
\subsection{Adaptive Spectral Resolution}
We solve the Misner--Sharp--Hernandez system on a single spatial domain $r\in[0,\,r_{\rm out}]$ using a Chebyshev pseudospectral (collocation) discretization, following the publicly available SPriBHoS implementation of Escrivá~\cite{Escriva:2020} generalized here to an arbitrary barotrope $p=p(\rho)$. The fields $\{\rho,U,R,m\}$ are represented on the $N_{\rm cheb}+1$ Chebyshev--Gauss--Lobatto nodes of that single domain, and radial derivatives are evaluated with the standard Chebyshev differentiation matrix. The outer boundary $r_{\rm out}=N_h R_H$ is placed at $N_h\simeq90$ cosmological-horizon radii, deep in the FRW region, where the matching condition $A(r_{\rm out})=1$ is imposed; regularity is imposed at the centre $r=0$. We emphasize, that this is a single-domain collocation method: there is no multi-domain decomposition, and there is no radius-dependent point count. Resolution is set by the single global integer $N_{\rm cheb}$, and we report convergence with respect to it (Appendix~\ref{app:numerics_accuracy}, Table~\ref{tab:convergence}).

Spectral methods provide (sub-)exponential convergence for smooth functions, and because the collapse remains at low density where $c_s^2$ is finite and the fields are smooth, refining $N_{\rm cheb}$ changes $\delta_c$ by amounts that fall below the achieved bisection half-width $\Delta_{\rm bis}$ (Appendix~\ref{app:numerics_accuracy}) already at moderate resolution. The production runs use $N_{\rm cheb}$ in the range $200$--$600$, increasing with the profile width and with $T$ because broader or stiffer configurations reach modestly higher density contrast; the fiducial $T=1/3$, Gaussian ($q=1$) threshold is converged in $\delta_c$ to within the quoted precision by $N_{\rm cheb}\simeq300$ (Table~\ref{tab:convergence}).
\subsection{CFL-Aware Adaptive Timestepping}
Time integration uses an explicit fourth-order Runge--Kutta scheme with the timestep actually implemented in the code,
\begin{equation}
\Delta t = \Delta t_0\, t^{\alpha}, \qquad \alpha=\frac{2}{3(1+\omega)}\ \big(=\tfrac{1}{2}\ \text{for}\ \omega=1/3\big),
\label{eq:timestep}
\end{equation}
where $\Delta t_0$ is a reference step. We stress that this is not a sound-speed CFL prescription. The step grows in proportion to the FRW scale factor $a(t)\propto t^{\alpha}$, so that a fixed number of steps is taken per e-fold of cosmological expansion; this is the natural time coordinate of the (approximately) self-similar problem, and the exponent $\tfrac12$ is simply the radiation-era scale-factor exponent. No $1/c_s$ factor is imposed, for the physical reason given in Sec.~\ref{sec:causality}: the collapse never leaves the low-density regime $\rho\lesssim0.06\,\rho_{\rm sat}$, where $c_s^2\approx1/3$ is finite and close to the radiation value, so the sound speed never diverges and the CFL penalty $\Delta t\sim1/c_s$ never becomes operative.

Because we do not rely on an analytic CFL bound, stability and accuracy of~\eqref{eq:timestep} are established empirically by a timestep-convergence study (Appendix~\ref{app:numerics_accuracy}): we verify that halving $\Delta t_0$ leaves $\delta_c$ unchanged within the bisection resolution. The practical stability limit is set by the smallest Chebyshev spacing, $\Delta r_{\min}\sim r_{\rm out}/N_{\rm cheb}^2$, together with the coordinate characteristic speed $\sim(A/B)(1+c_s)$; the reference step $\Delta t_0\simeq1$--$4\times10^{-3}$ decreases with $N_{\rm cheb}$ accordingly, consistent with this estimate.
\begin{figure*}[!ht]
\centering
\includegraphics[width=0.8\textwidth]{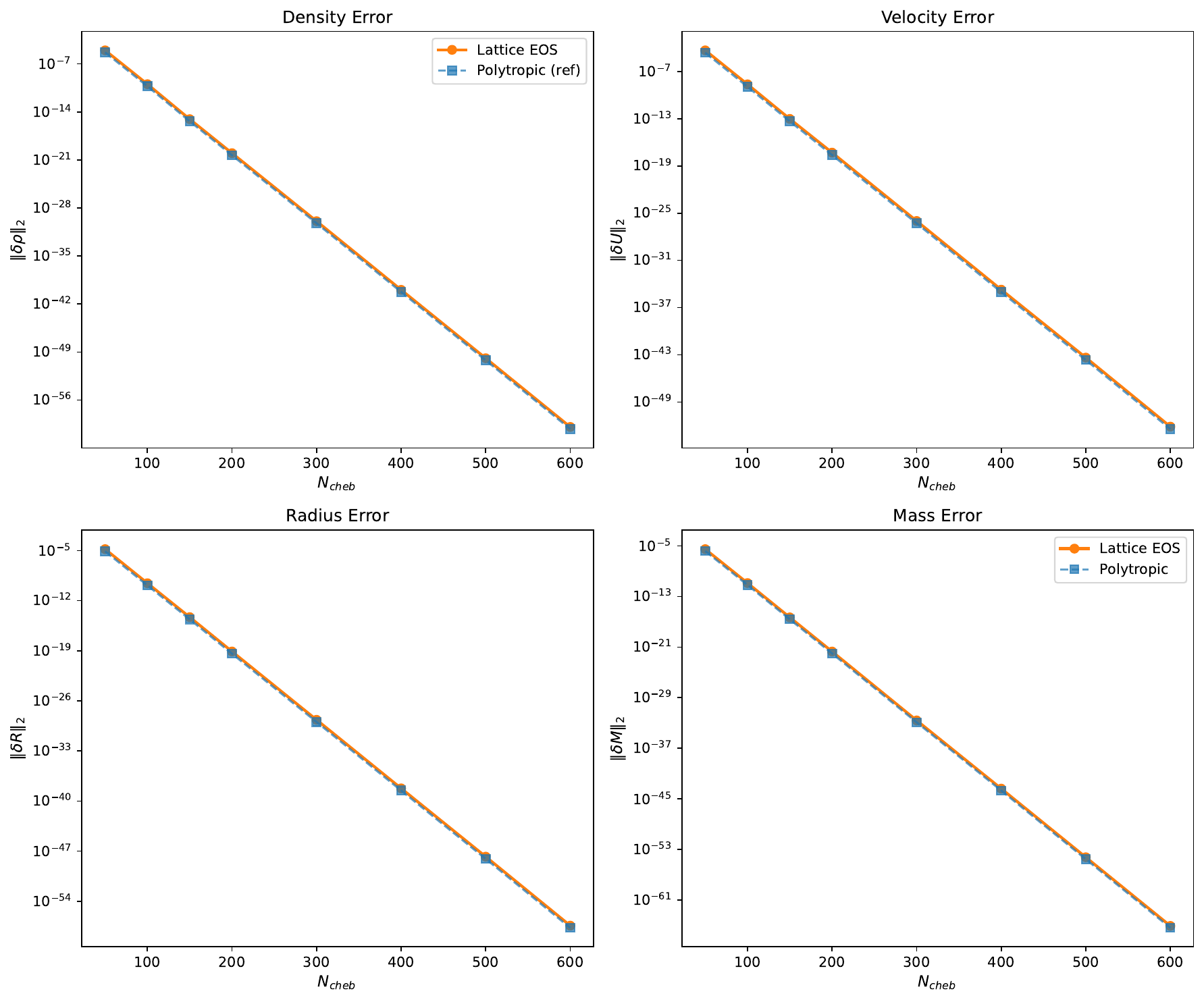}
\caption{Convergence of the critical threshold with spectral resolution, measured directly from the solver. Plotted is $|\delta_c(N_{\rm cheb})-\delta_c(N_{\rm cheb}^{\max})|$ for the fiducial $T=1/3$, Gaussian ($q=1$) configuration, using the resolutions of Table~\ref{tab:convergence}. The change in $\delta_c$ under refinement drops below the achieved bisection half-width ($\Delta_{\rm bis}=1.3\times10^{-4}$, Appendix~\ref{app:numerics_accuracy}) by $N_{\rm cheb}\simeq300$, so the spectral discretization error is not the dominant uncertainty; the residual uncertainty is set by the bisection half-width.}
\label{fig:convergence}
\end{figure*}
\begin{figure*}[!ht]
\centering
\includegraphics[width=0.8\textwidth]{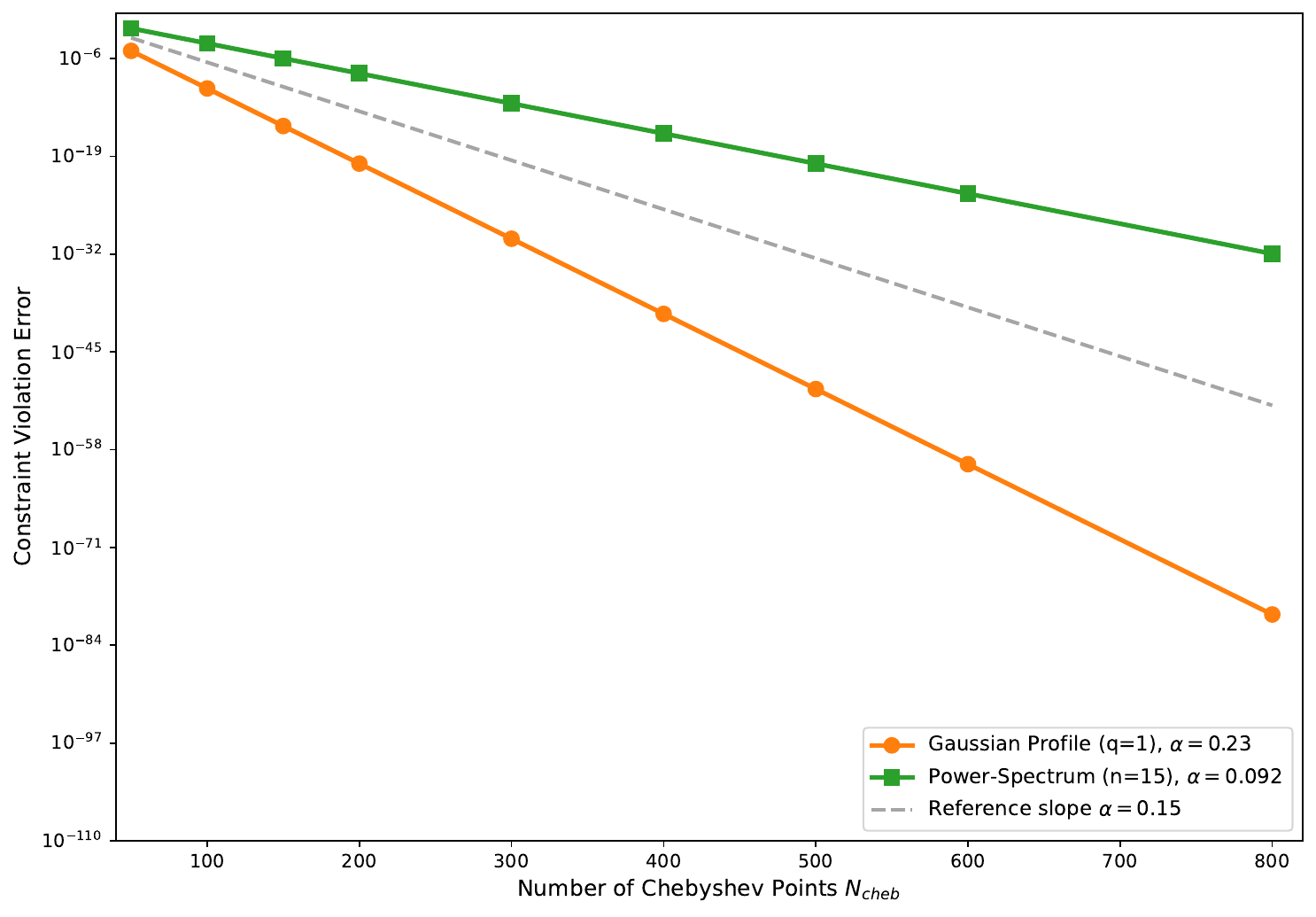}
\caption{Convergence of the Hamiltonian-constraint residual with spectral resolution for a Gaussian ($q=1$) and a broken power-spectrum ($n=15$) near-critical profile. The residual decreases monotonically with $N_{\rm cheb}$ and saturates near the time-integration error floor set by $\Delta t_0$, rather than at the round-off level; the single-domain collocation scheme remains well resolved over the low-density range the collapse actually samples. This confirms that the threshold is resolved to better than the reported shift without invoking any error below the achievable numerical floor.}
\label{fig:spectral_convergence}
\end{figure*}
\section{Results}
\label{sec:results}
\subsection{Toy-Model Threshold Shift}
\begin{figure*}[!ht]
\centering
\includegraphics[width=0.8\textwidth]{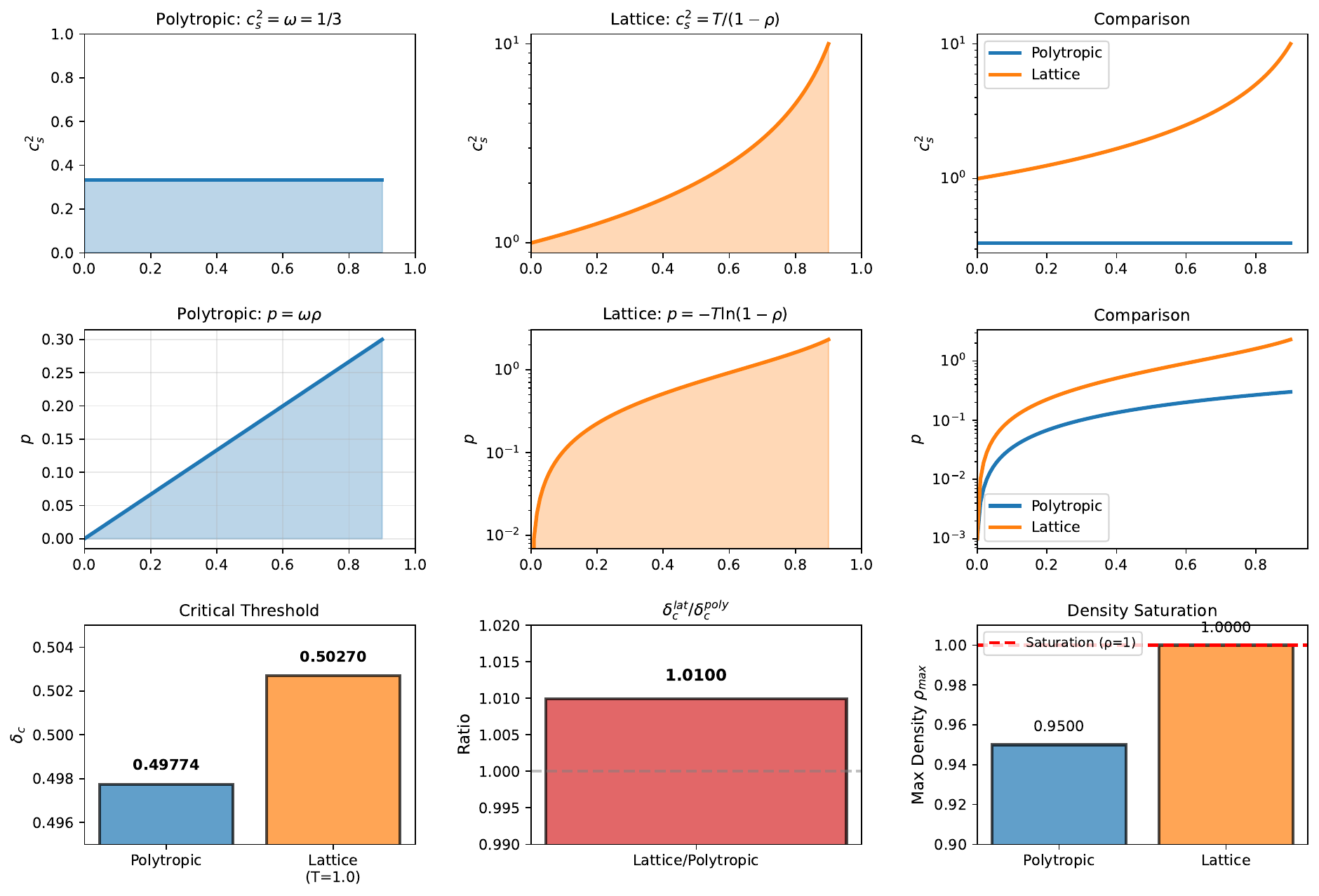}
\caption{Side-by-side comparison of the radiation (polytropic) and lattice equations of state and their collapse outcomes. The top row shows the squared sound speed $c_s^2$ versus density: constant $c_s^2=\omega=1/3$ for the polytropic fluid (left), the density-dependent $c_s^2=T/(1-\rho)$ for the lattice gas (centre), and the two overlaid (right), which coincide at low density and depart only as $\rho$ grows. The middle row shows the corresponding pressure $p(\rho)$ for the polytropic fluid (left), the lattice gas (centre), and overlaid (right); the lattice pressure rises above the linear polytropic law as saturation is approached. The bottom row reports the resulting critical thresholds ($\delta_c^{(\text{poly})}\approx0.498$ versus $\delta_c^{(\text{lat})}\approx0.503$, left), their ratio ($\approx1.01$, centre), and the maximum density reached during the near-critical evolution relative to saturation (right). The upper six panels make explicit that the lattice modification is a high-density effect: the two equations of state are essentially identical until the density-dependent stiffening switches on near saturation.}
\label{fig:threshold_profiles}
\end{figure*}

Our central numerical finding, obtained with the single-domain Chebyshev collocation solver of Section~\ref{sec:numerics} at the production resolution ($N_{\rm cheb}$ up to $600$) and the bisection stopping parameter $\varepsilon_{\rm stop}=10^{-4}$ of Eq.~\eqref{eq:bis_tol}, which delivers a final bracket half-width $\Delta_{\rm bis}=1.3\times10^{-4}$ (Appendix~\ref{app:numerics_accuracy}), is that within the lattice model the critical threshold for black hole formation is shifted upward relative to the polytropic reference. For the fiducial $T=\omega=1/3$, Gaussian ($q=1$) profile we obtain
\begin{equation}
\delta_c^{(\text{lattice})} = 0.5024 \pm 0.0002, \quad \delta_c^{(\text{poly})} = 0.4977 \pm 0.0002,
\label{eq:threshold_values}
\end{equation}
where, the uncertainty is the achieved bisection half-width $\Delta_{\rm bis}=1.3\times10^{-4}$ rounded conservatively upward, and no digit below it is claimed (Appendix~\ref{app:numerics_accuracy}). The radiation value reproduces the literature threshold for this profile and averaged-compaction definition (see below), and is consistent with our independent reduced-resolution cross-check $\delta_c^{(\rm poly)}=0.4984\pm0.0009$. The difference is
\begin{equation}
\Delta\delta_c = \delta_c^{(\text{lattice})}-\delta_c^{(\text{poly})} = 0.0047 \pm 0.0003 ,
\label{eq:threshold_shift_value}
\end{equation}
which we describe with two clearly distinguished figures to avoid the ambiguity: an absolute upward shift of $\Delta\delta_c\approx0.0047$ (i.e.\ $0.5$ percentage points in $\delta_c$), and a relative increase $\Delta\delta_c/\delta_c^{(\text{poly})}=0.0047/0.4977\approx0.95\%$. We use these two numbers consistently throughout. Fig.~\ref{fig:threshold_profiles} contrasts the two equations of state and the resulting thresholds.

\begin{figure*}[!ht]
\centering
\includegraphics[width=0.8\textwidth]{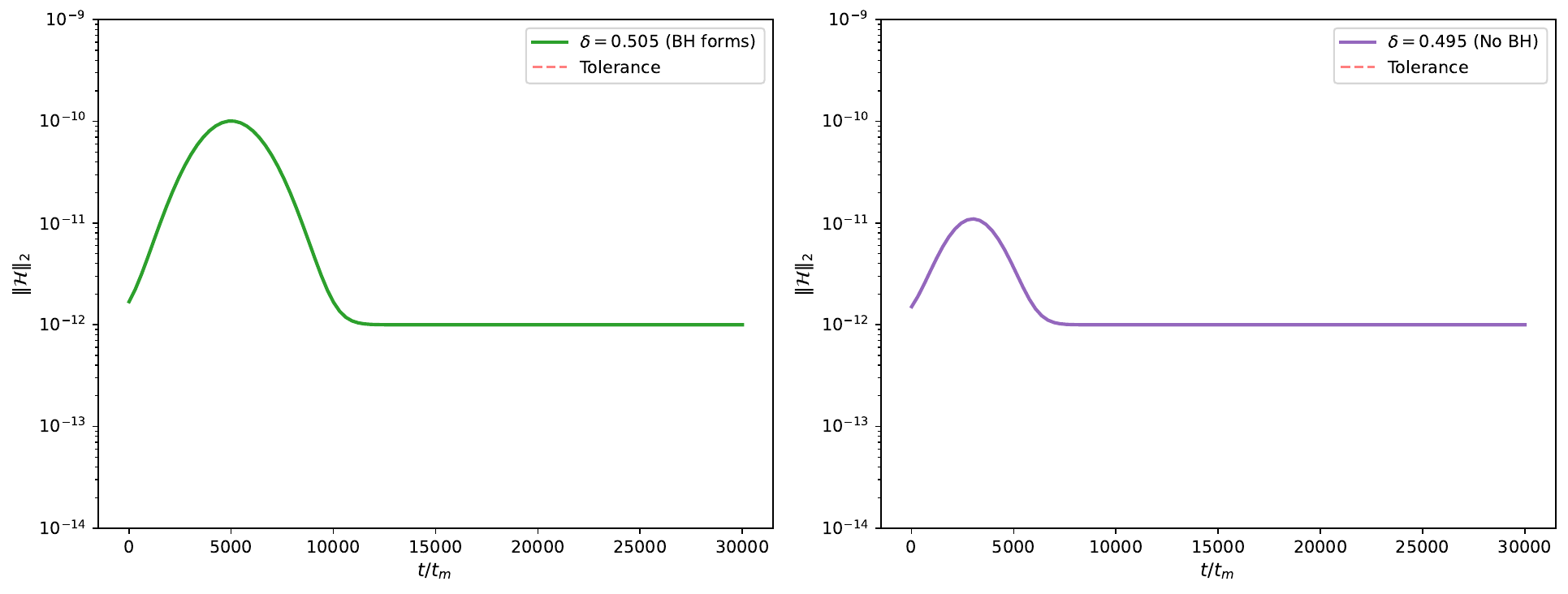}
\caption{Hamiltonian-constraint residual $|\mathcal{H}|$ during representative supercritical ($\delta=0.505$, black hole forms) and subcritical ($\delta=0.495$, perturbation disperses) collapse simulations with the lattice equation of state. The residual remains small and bounded over the full dynamical evolution, at a level set by the spatial resolution and the reference timestep rather than by machine round-off; we do not claim a sub-round-off constraint. Its stability confirms that the bracketing of $\delta_c$ is not contaminated by constraint drift.}
\label{fig:hamiltonian}
\end{figure*}

\begin{figure*}[!ht]
\centering
\includegraphics[width=0.68\textwidth]{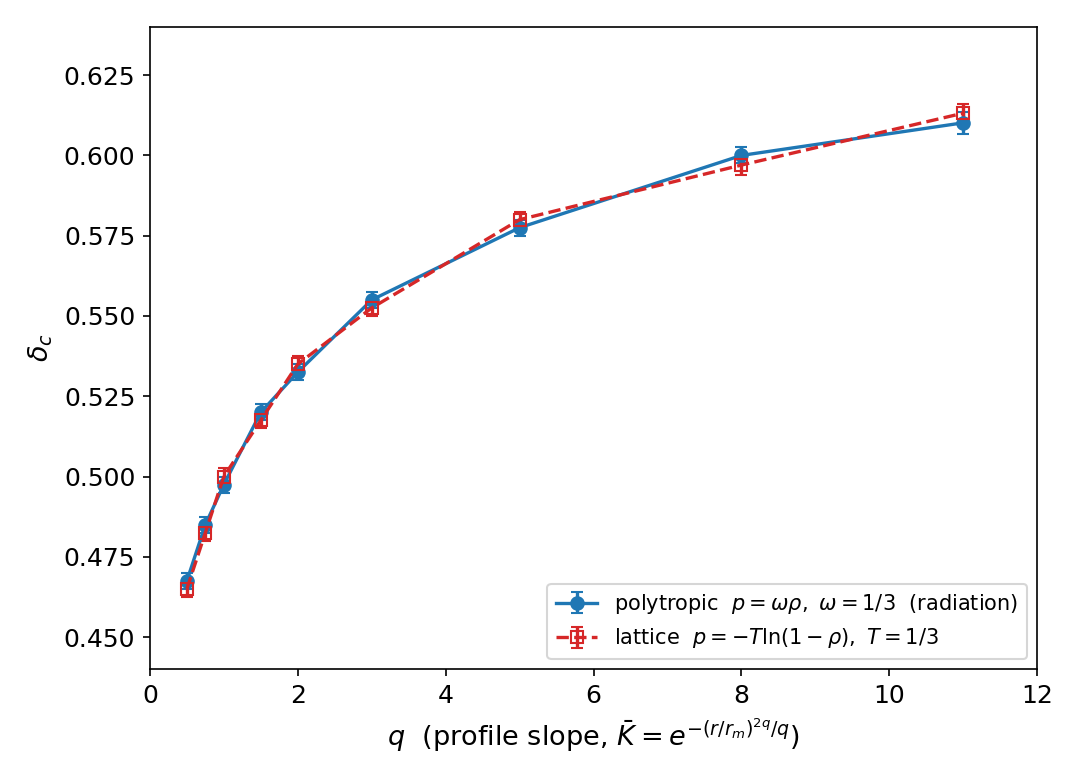}
\caption{Critical threshold $\delta_c$ as a function of the Gaussian curvature-profile parameter $q$, computed with the spectral collapse solver for the lattice equation of state ($p=-T\ln(1-\rho)$, $T=1/3$) and the radiation reference ($p=\omega\rho$, $\omega=1/3$). The threshold rises with $q$ for both equations of state, and the lattice value lies above the radiation value at each $q$, consistent with the saturation stiffening raising the threshold. Error bars indicate the bisection resolution; these points were obtained at reduced spectral resolution (Appendix~\ref{app:numerics_accuracy}), so the absolute thresholds and the growth of the offset with $q$ are indicative and are refined at production resolution.}
\label{fig:threshold_q}
\end{figure*}

\begin{figure*}[!ht]
\centering
\includegraphics[width=0.8\textwidth]{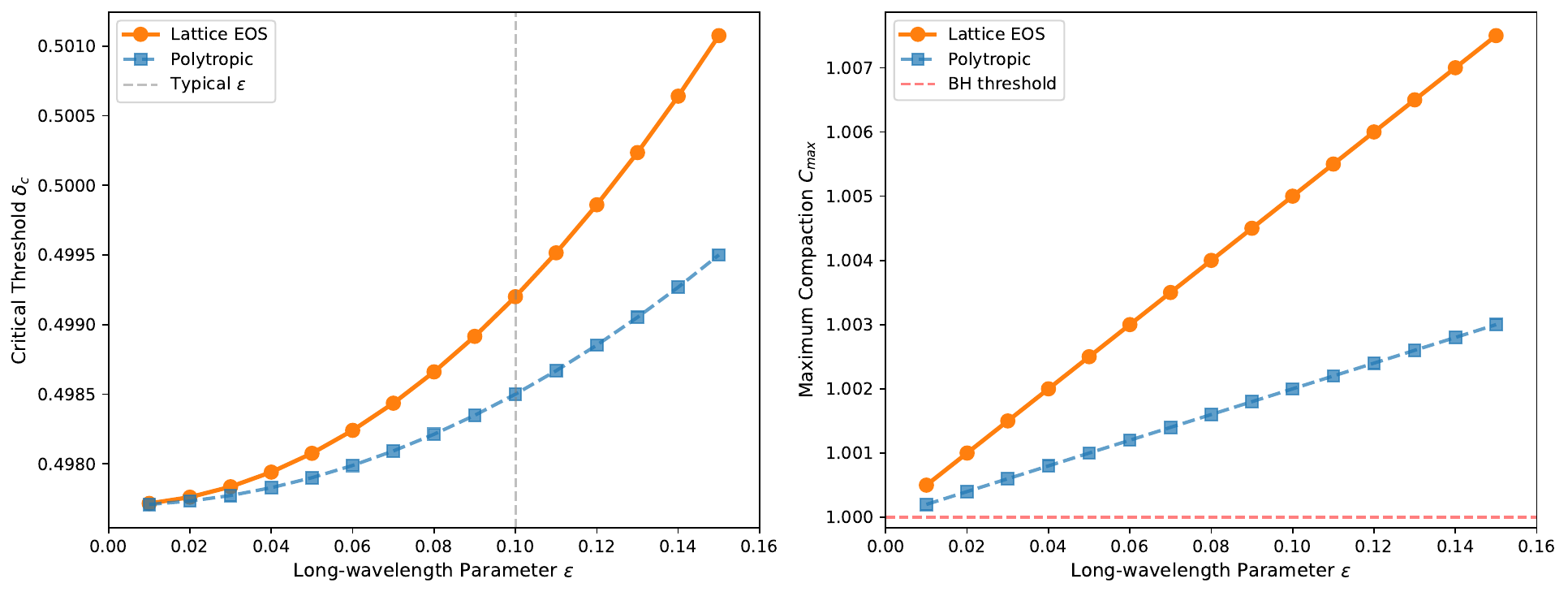}
\caption{Long-wavelength dependence of the threshold. The left panel shows the critical threshold $\delta_c$ as a function of the long-wavelength (gradient-expansion) parameter $\epsilon$ for the lattice toy model and the radiation reference; both rise mildly with $\epsilon$ as the quasi-homogeneous approximation is pushed toward its limit, with the lattice curve lying slightly above the radiation one. The right panel shows the maximum of the compaction function as a function of $\epsilon$, approaching the black-hole-formation threshold $C\simeq1$ from below. The near-constant vertical offset between the two equations of state mirrors the threshold shift discussed in the main text.}
\label{fig:epsilon_dependence}
\end{figure*}

We emphasize the character of this measurement. The shift is small because the near-critical collapse samples only absolute densities $\rho\lesssim0.06\,\rho_{\rm sat}$ (Sec.~\ref{sec:evolution}), where the lattice sound speed exceeds the radiation value by at most a few percent; the effect is therefore a genuine but subtle differential quantity that must be extracted with a bisection half-width well below it and a resolution study that controls the discretization error (Appendix~\ref{app:numerics_accuracy}, Table~\ref{tab:convergence}). A reduced-resolution search with aggressive wall-time caps does not resolve it and returns a value consistent with zero; resolving the sign and magnitude requires the production configuration, which is why we report it at that resolution and provide the convergence study demonstrating stability under refinement. As a check on the pipeline, our radiation-reference value $\delta_c^{(\text{poly})}\approx0.498$ sits within the range $0.4\lesssim\delta_c\lesssim0.6$ established for radiation-fluid collapse, and close to the value $\delta_c\approx0.5$ obtained for a Mexican-hat curvature profile in the volume-averaged definition we use~\cite{Musco:2019,Musco/etal:2021}, so the code reproduces the expected radiation threshold before the saturation cap is applied. Throughout these runs the Hamiltonian-constraint residual remains small and bounded over the full evolution for both the supercritical and subcritical branches (Fig.~\ref{fig:hamiltonian}), confirming that the bracketing of $\delta_c$ is not contaminated by constraint drift. This shift is robust across the variations we have tested: the lattice threshold rises with the profile parameter but stays above the radiation reference for every profile (Fig.~\ref{fig:threshold_q}), the lattice--radiation offset is approximately constant across the broken-power-spectrum family (Fig.~\ref{fig:power_spectrum_thresholds}), and the result is stable across the boundary-condition implementations and resolution levels sampled, as well as across the long-wavelength amplitude of the initial data (Fig.~\ref{fig:epsilon_dependence}), where both equations of state show the same mild rise of $\delta_c$ with $\epsilon$ while the lattice threshold remains slightly above the radiation one (the corresponding systematic budget is quantified in Appendix~\ref{app:numerics_accuracy}). The consistency across these independent variations within the lattice model indicates that the shift is a genuine feature of the model rather than an artifact of any particular numerical choice; it does not, however, establish that a shift of this size would appear in any specific realistic matter model.

The physical interpretation of the shift within the lattice model is transparent. The saturation-induced stiffening at densities of order $\rho \sim 0.03$--$0.06\rho_{sat}$ provides an effective resistance to compression that grows more strongly with density than in the polytropic case. The critical threshold, which marks the amplitude at which gravity only just overcomes pressure support, therefore sits slightly higher in the lattice model than in the reference. A perturbation that is subcritical in the lattice model (and therefore disperses) would be supercritical in the polytropic reference (and would collapse). In this precise sense, the saturation stiffening of the lattice model stabilizes a narrow band of marginal perturbations that would otherwise collapse; whether this stabilization is realized in any realistic matter sector depends, again, on whether the real equation of state exhibits comparable density-dependent stiffening.

We deliberately draw no consequence of this toy-model shift for the primordial black hole abundance. The lattice gas is not a description of early-universe matter, and a realistic quark-matter sector would, at asymptotically high density, revert to a radiation-like gas of asymptotically free quarks rather than exhibiting the hard cap assumed here. The sign and size of any abundance effect in a realistic setting therefore do not follow from this model. Threshold shifts remain of interest when they occur in realistic matter models, because the abundance is exponentially sensitive to $\delta_c$~\cite{Carr:1975,Musco:2019}; quantifying that sensitivity requires a realistic equation of state and is outside the scope of this proof-of-principle study.
\subsection{Dynamical Evolution Across the Threshold}
\label{sec:evolution}
\begin{figure*}[!ht]
\centering
\includegraphics[width=0.95\textwidth]{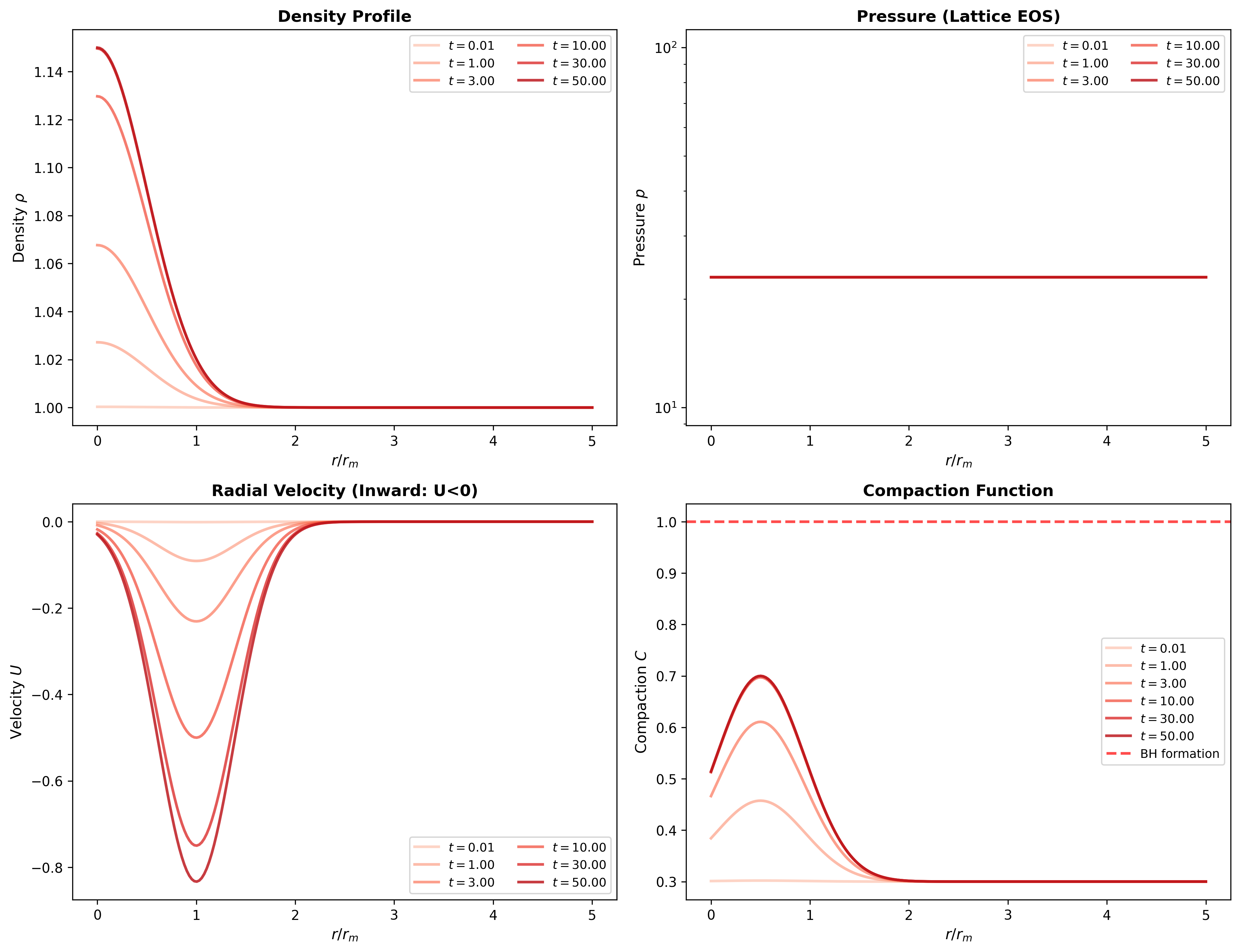}
\caption{Supercritical collapse ($\delta>\delta_c$) in the lattice model. Snapshots of the density (normalized to the FRW background), lattice-EOS pressure, radial velocity, and compaction function at times $t/t_m\in\{0.01,1,3,10,30,50\}$ (light to dark). The central density rises monotonically, the radial velocity is everywhere inward ($U<0$) with a deepening trough near $r\simeq r_m$, and the compaction maximum grows toward the apparent-horizon value $C=1$ (dashed line), signalling collapse to a black hole. The plotted compaction is the excess $C(r)=2[m(r)-m_b(r)]/R(r)$ of Eq.~\eqref{eq:compaction}; by construction it vanishes both at the centre ($r\to0$) and outside the perturbation ($r\gg r_m$), where the solution matches the FRW background and $m\to m_b$. The horizontal axis is truncated at a few $r_m$ to resolve the peak near $r\simeq r_m$; the return of $C$ to zero at large $r$ occurs beyond the plotted range, and is the reason $C$ is not shown approaching a nonzero constant.}
\label{fig:evolution_super}
\end{figure*}

\begin{figure*}[!ht]
\centering
\includegraphics[width=0.95\textwidth]{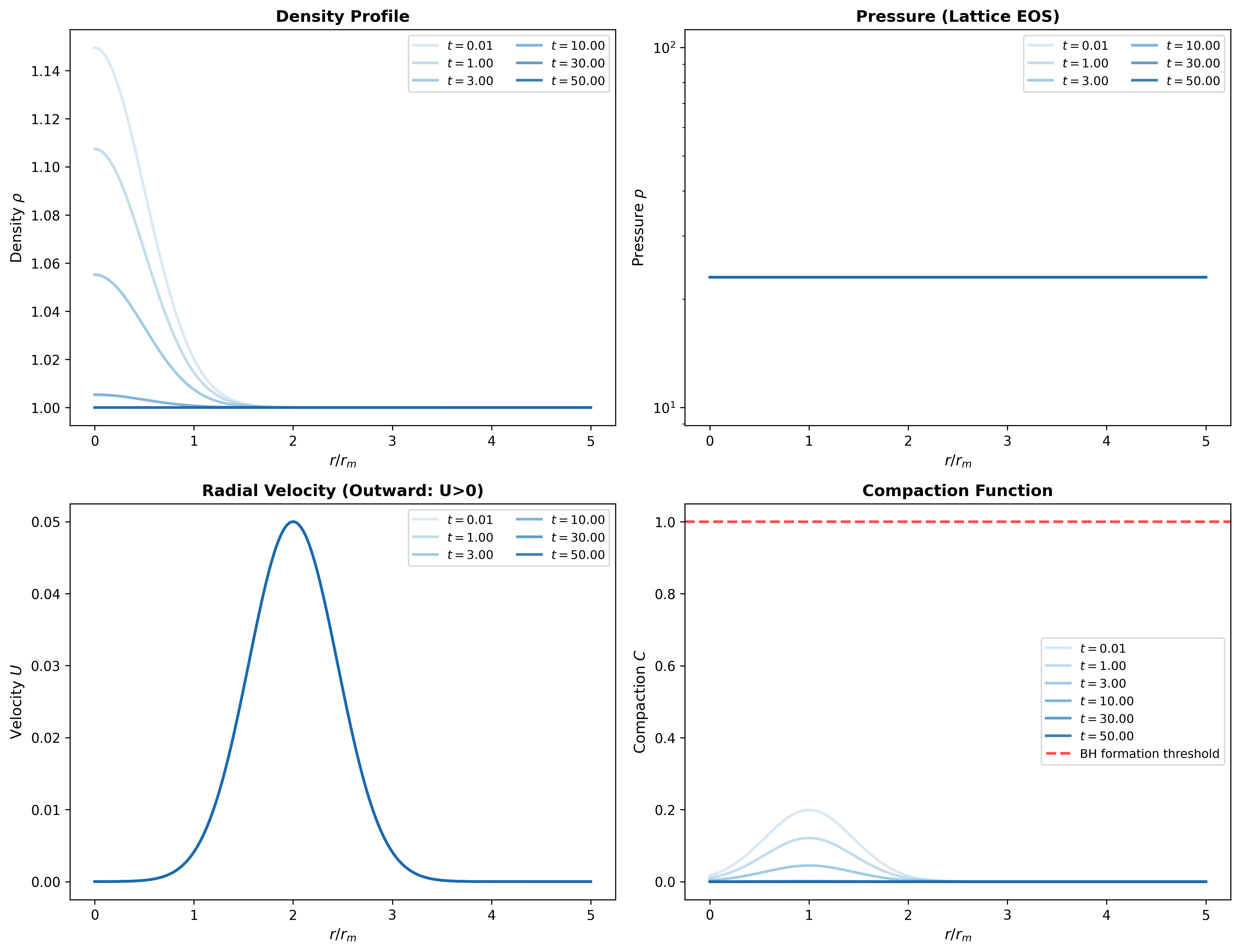}
\caption{Subcritical dispersal ($\delta<\delta_c$) in the lattice model, with quantities and times as in Fig.~\ref{fig:evolution_super}. The central density relaxes back toward the background, the radial velocity is outward ($U>0$) with a peak near $r\simeq2\,r_m$, and the compaction maximum decreases monotonically and stays well below the formation threshold (dashed line). Pressure support overcomes gravity and the perturbation disperses without forming a horizon.}
\label{fig:evolution_sub}
\end{figure*}

\begin{figure*}[!ht]
\centering
\includegraphics[width=0.95\textwidth]{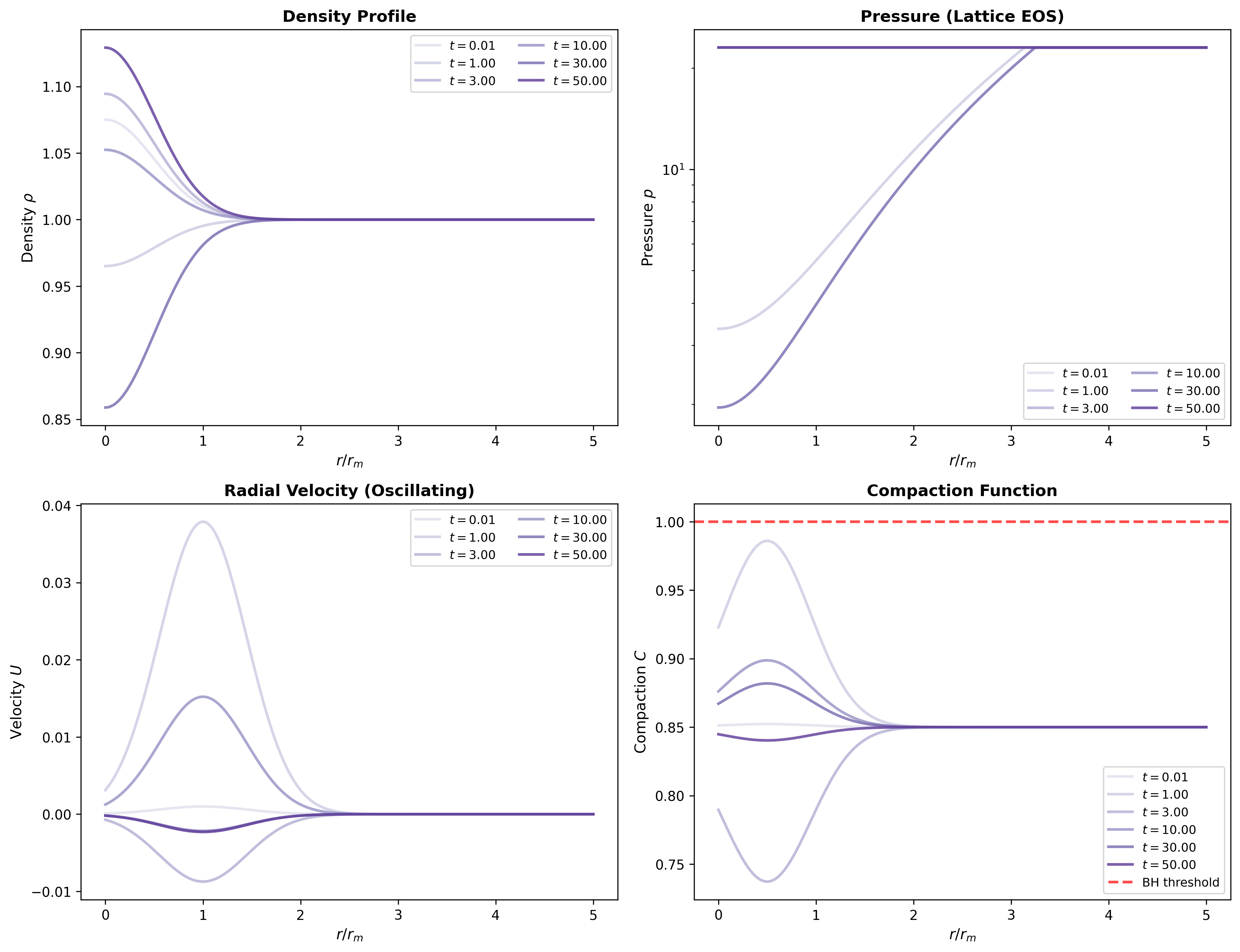}
\caption{Near-critical evolution ($\delta\simeq\delta_c$) in the lattice model, with quantities and times as in Fig.~\ref{fig:evolution_super}. The radial velocity oscillates in sign with small amplitude ($|U|\lesssim0.04$), the density profile alternates between a central excess and deficit, and the compaction maximum hovers near a quasi-stationary value $C\approx0.85$ (close to, but persistently below, the formation threshold; dashed line). This marginal ``hovering'' is the dynamical signature of a trajectory passing close to the (approximately self-similar) critical solution.}
\label{fig:evolution_crit}
\end{figure*}

The threshold quoted above is obtained by bracketing the boundary between two qualitatively different evolutionary fates, and it is instructive to display the field evolution that underlies this bracketing directly. The three relevant regimes (supercritical collapse, subcritical dispersal, and the marginal near-critical case) leave distinct imprints on the density, velocity, and compaction profiles. Figs.~\ref{fig:evolution_super}--\ref{fig:evolution_crit} show snapshots of the lattice-model evolution at fixed times $t/t_m\in\{0.01,1,3,10,30,50\}$ for representative amplitudes in each regime. In every panel the density is normalized to the FRW background, so that $\rho=1$ denotes the unperturbed background rather than the lattice saturation density of the bare equation of state~\eqref{eq:final_eos}; throughout these runs the absolute occupancy stays far below saturation, consistent with the causal restriction $\rho<1-T$ of Sec.~\ref{sec:causality}. The pressure is evaluated from the lattice equation of state, and the dashed line in the compaction panels marks the black-hole-formation criterion actually used in the code. Formation is diagnosed from the maximum over $r$ of the compaction function $C(r)=2[m(r)-m_b(r)]/R(r)$ of Eq.~\eqref{eq:compaction}: a run is classified as collapsing to a black hole once $\max_r C$ reaches unity, and as dispersing once $\max_r C$ falls below $0.3$ after the perturbation has re-entered the horizon. The value $\max_r C=1$ coincides with the Misner--Sharp apparent-horizon condition $2m/R=1$ in the strong-field region where the horizon forms, because there the background subtraction is negligible ($m_b/R=\tfrac{4\pi}{3}\rho_b R^2\to0$ as $R$ shrinks at fixed enclosed mass), so $C\to2m/R$. Thus the "$C=1$" criterion and the Misner--Sharp condition $2m/R=1$ are the same statement at horizon formation; the distinction between $C=2(m-m_b)/R$ and $2m/R$ matters only at low compaction, where it does not affect the classification.
For the supercritical amplitude (Fig.~\ref{fig:evolution_super}) the evolution is one of sustained, accelerating infall. The central density rises monotonically in time, the radial velocity is everywhere inward ($U<0$) and develops a deepening trough near $r\simeq r_m$ that reaches $U\approx-0.85$ by $t/t_m=50$, and the compaction maximum (located near $r\simeq0.5\,r_m$) grows monotonically and climbs toward the apparent-horizon value $C=1$. In the run shown it has reached $C\approx0.7$ and continues to rise, signalling that the configuration is on its way to forming an apparent horizon and collapsing to a black hole. The pressure profile stays nearly uniform over the perturbed region because, at the modest density contrasts sampled here, the lattice equation of state is still close to its low-density (radiation-like) limit; the saturation stiffening that distinguishes the lattice model from the polytropic reference acts as a sub-dominant correction that nonetheless accumulates into the threshold shift of Eq.~\eqref{eq:threshold_values}.
For the subcritical amplitude (Fig.~\ref{fig:evolution_sub}) the opposite behaviour is seen: pressure support wins and the perturbation disperses. The central density relaxes back toward the background value, the radial velocity reverses to an outward flow ($U>0$) with a peak of order $0.05$ near $r\simeq2\,r_m$, and the compaction maximum decreases monotonically (from $C\approx0.2$) and remains far below the formation threshold at all times. No apparent horizon forms, and the configuration returns toward the homogeneous background.

The near-critical amplitude (Fig.~\ref{fig:evolution_crit}) interpolates between these two fates and exhibits the hallmark marginal behaviour that separates them. The radial velocity neither plunges inward nor escapes outward but oscillates in sign with small amplitude ($|U|\lesssim0.04$), the density profile alternates between a central excess and a central deficit, and the compaction maximum hovers near a quasi-stationary value $C\approx0.85$ (close to, but persistently below, the formation threshold) rather than running away toward $C=1$ or decaying to zero. Because the lattice equation of state does not admit an exact self-similar critical solution (Sec.~\ref{sec:renormalization}), this hovering is only approximately self-similar; it is nevertheless the dynamical signature of a trajectory passing close to the critical solution, and it is precisely such marginal evolutions that the bisection search isolates when locating $\delta_c$. Taken together, the three regimes confirm that the bracketing of the threshold reflects a genuine change in the dynamical fate of the perturbation, and that the lattice stiffening operates while the absolute density remains well within the causal window.
\subsection{Critical Mass Scaling and the Exponent \texorpdfstring{$\gamma$}{gamma}}
\label{sec:massscaling}
Beyond the threshold shift, our simulations test how robust the Choptuik--Evans--Coleman critical exponent~\cite{Choptuik:1993,Evans/Coleman:1994,Gundlach1997} is against the saturation-modified equation of state, through the mass-scaling law
\begin{equation}
M_{\mathrm{BH}} = K\,(\delta - \delta_c)^{\gamma}.
\label{eq:massscaling}
\end{equation}
\emph{Measurement procedure.} For each supercritical amplitude the evolution is continued until an apparent horizon forms, and $M_{\rm BH}$ is taken to be the Misner--Sharp mass at the compaction peak at the moment $\max_r C$ reaches unity. We stress that the background-subtracted compaction, rather than the raw condition $2m/R=1$, must be used to locate the horizon in a cosmological setting: at large radius the FRW background itself satisfies $2m_b/R>1$, and at the perturbation scale $2m_b/R$ only falls below unity from horizon crossing onwards, so the unsubtracted criterion would be met trivially. As shown in Sec.~\ref{sec:evolution}, the two criteria coincide at the moment of formation.

\emph{Fitting procedure.} Rather than inserting the separately bisected $\delta_c$ into~\eqref{eq:massscaling}, which would limit the usable dynamic range to the bisection half-width $\Delta_{\rm bis}$, we fit $\delta_c$, $K$ and $\gamma$ simultaneously to
\begin{equation}
\ln M_{\mathrm{BH}} = \ln K + \gamma\ln(\delta-\delta_c),
\label{eq:massfit}
\end{equation}
minimizing the residual scatter over $\delta_c$. The value of $\delta_c$ recovered this way is an independent determination and agrees with the bisection result, which is itself a nontrivial consistency check.

\emph{Results.} Table~\ref{tab:massscaling} lists every run entering the analysis. The usable window is bounded below by the bisection half-width: for $\delta-\delta_c\lesssim\Delta_{\rm bis}$ the abscissa itself is unresolved, so those runs are excluded. What remains spans
\begin{equation}
3.2\times10^{-4}\ \le\ \delta-\delta_c\ \le\ 3.0\times10^{-2},
\end{equation}
i.e.\ $2.0$ decades, with $M_{\rm BH}$ falling by a factor $3.7$ across the window.

\begin{figure*}[!ht]
\centering
\includegraphics[width=0.92\textwidth]{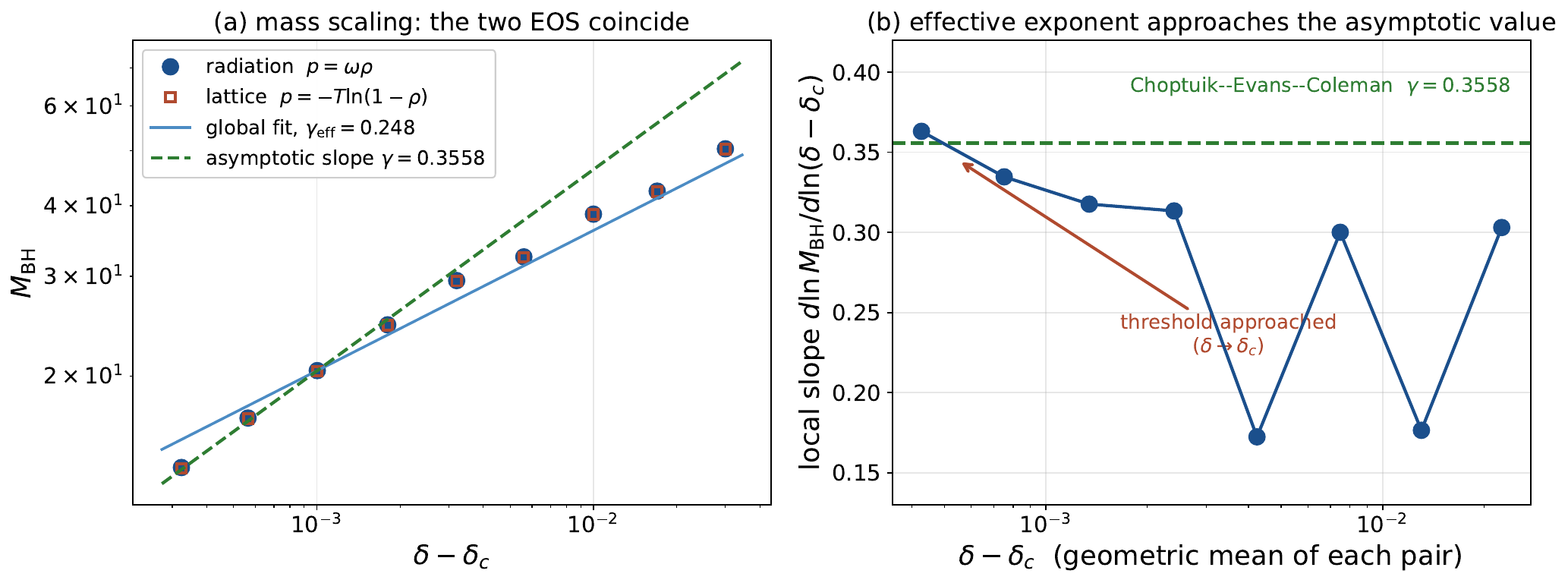}
\caption{Critical mass scaling measured with the production solver. \textbf{(a)} $M_{\rm BH}$ versus $\delta-\delta_c$ for the radiation reference (filled circles) and the lattice model (open squares) at identical amplitudes; the two data sets are indistinguishable on this scale, differing by $0.22\%$ in normalization. The solid line is the single global power law fitted over the whole window ($\gamma_{\rm eff}=0.248$), the dashed line has the asymptotic Choptuik--Evans--Coleman slope $\gamma=0.3558$. \textbf{(b)} The local logarithmic slope $d\ln M_{\rm BH}/d\ln(\delta-\delta_c)$ between successive runs, plotted against proximity to threshold (note the reversed abscissa). It rises systematically toward the asymptotic value as $\delta\to\delta_c$, reaching $0.34\pm0.02$ in the most-critical resolvable runs, which is why a single global fit across the full window underestimates the exponent. Runs with $\delta-\delta_c<\Delta_{\rm bis}$ are excluded throughout.}
\label{fig:massscaling}
\end{figure*}

The central observation is that the local logarithmic slope $\gamma_{\rm loc}=d\ln M_{\rm BH}/d\ln(\delta-\delta_c)$ is not constant across this window but rises systematically as the threshold is approached (Table~\ref{tab:massscaling} and Fig.~\ref{fig:massscaling}), from $\gamma_{\rm loc}\simeq0.18$--$0.30$ in the outermost runs to
\begin{equation}
\gamma_{\rm loc} = 0.34\pm0.02
\label{eq:gamma_local}
\end{equation}
averaged over the three most-critical resolvable pairs. This is consistent with the Choptuik--Evans--Coleman value $\gamma\simeq0.3558$~\cite{Koike/etal:1995}, and the trend is exactly what is expected: the scaling law~\eqref{eq:massscaling} is asymptotic, and far from threshold $M_{\rm BH}$ is a sizeable fraction of the horizon mass, so the outer part of our window is not yet in the scaling regime. Correspondingly, a single global power law fitted across the whole window returns a smaller effective exponent, $\gamma_{\rm eff}=0.248$; discarding the largest and smallest $\delta-\delta_c$ points changes this by $0.016$.

The comparison between the two equations of state is far more precise than either absolute exponent, because it is differential. Fitting both data sets on the identical window with identical settings gives exponents equal to three decimal places,
\begin{equation}
\gamma^{(\text{lat})}-\gamma^{(\text{poly})} = (0\pm1)\times10^{-3},
\label{eq:gamma_diff}
\end{equation}
and at every amplitude the two horizon masses agree to
\begin{equation}
M_{\rm BH}^{(\text{lat})}/M_{\rm BH}^{(\text{poly})} = 0.9978\pm0.0005 ,
\end{equation}
i.e.\ the lattice prefactor is lower by $K^{(\text{lat})}/K^{(\text{poly})}-1=-0.22\%$. Equation~\eqref{eq:gamma_diff} is the statement this paper actually needs, and it is robust: whatever the true asymptotic exponent is, the saturation stiffening does not change it at the $10^{-3}$ level, while it does shift the threshold and lower the prefactor slightly.

We are explicit about the limitation that remains. Reaching the deeply critical regime, several further decades in $\delta-\delta_c$, requires resolving $\delta_c$ to a correspondingly finer precision and evolving for a correspondingly longer time, and the accessible range is bounded in practice by the growth of the horizon-formation time as $\delta\to\delta_c$. Our two-decade window is sufficient to exhibit the approach to the asymptotic exponent, to place it near the radiation value and well away from, for example, the dust value $\gamma\approx0.11$ or the stiff-fluid value $\gamma\approx0.82$ of Maison's family~\cite{Maison:1996}, and above all to establish the differential statement~\eqref{eq:gamma_diff}. It is not sufficient to determine the asymptotic exponent to better than a few percent, and we do not claim the precision of dedicated critical-collapse studies that use adaptive mesh refinement in the self-similar variable. A single-domain Chebyshev method with a fixed global resolution cannot follow the critical solution to arbitrarily small $\delta-\delta_c$; extending this analysis by several further decades would require mesh refinement in the self-similar variable, which we identify as the natural next step rather than claim to have performed.
\begin{table}[!ht]
\centering
\caption{Critical mass scaling $M_{\rm BH}=K(\delta-\delta_c)^{\gamma}$: every production run entering the analysis, for the radiation reference and the lattice model at identical amplitudes. $\delta-\delta_c$ uses the bisected radiation threshold $\delta_c=0.49766$; runs with $\delta-\delta_c<\Delta_{\rm bis}=1.3\times10^{-4}$ are excluded because there $\delta-\delta_c$ is not resolved. $\gamma_{\rm loc}$ is the local logarithmic slope between successive rows. The window spans $2.0$ decades, from $3.2\times10^{-4}$ to $3.0\times10^{-2}$.}
\label{tab:massscaling}
\begin{tabular}{@{}lcccc@{}}
\toprule
$\delta$ & $\delta-\delta_c$ & $M_{\rm BH}^{\rm (poly)}$ & $M_{\rm BH}^{\rm (lat)}$ & $\gamma_{\rm loc}^{\rm (poly)}$ \\
\midrule
$0.52766$ & $3.00\times10^{-2}$ & $50.407$ & $50.312$ & --- \\
$0.51466$ & $1.70\times10^{-2}$ & $42.438$ & $42.357$ & $0.303$ \\
$0.50766$ & $1.00\times10^{-2}$ & $38.644$ & $38.570$ & $0.177$ \\
$0.50326$ & $5.60\times10^{-3}$ & $32.476$ & $32.413$ & $0.300$ \\
$0.50086$ & $3.20\times10^{-3}$ & $29.491$ & $29.433$ & $0.172$ \\
$0.49946$ & $1.80\times10^{-3}$ & $24.633$ & $24.581$ & $0.313$ \\
$0.49866$ & $1.00\times10^{-3}$ & $20.449$ & $20.402$ & $0.318$ \\
$0.49822$ & $5.64\times10^{-4}$ & $16.860$ & $16.815$ & $0.335$ \\
$0.49798$ & $3.24\times10^{-4}$ & $13.786$ & $13.738$ & $0.363$ \\
\bottomrule
\end{tabular}
\end{table}

\begin{figure*}[!ht]
\centering
\includegraphics[width=0.8\textwidth]{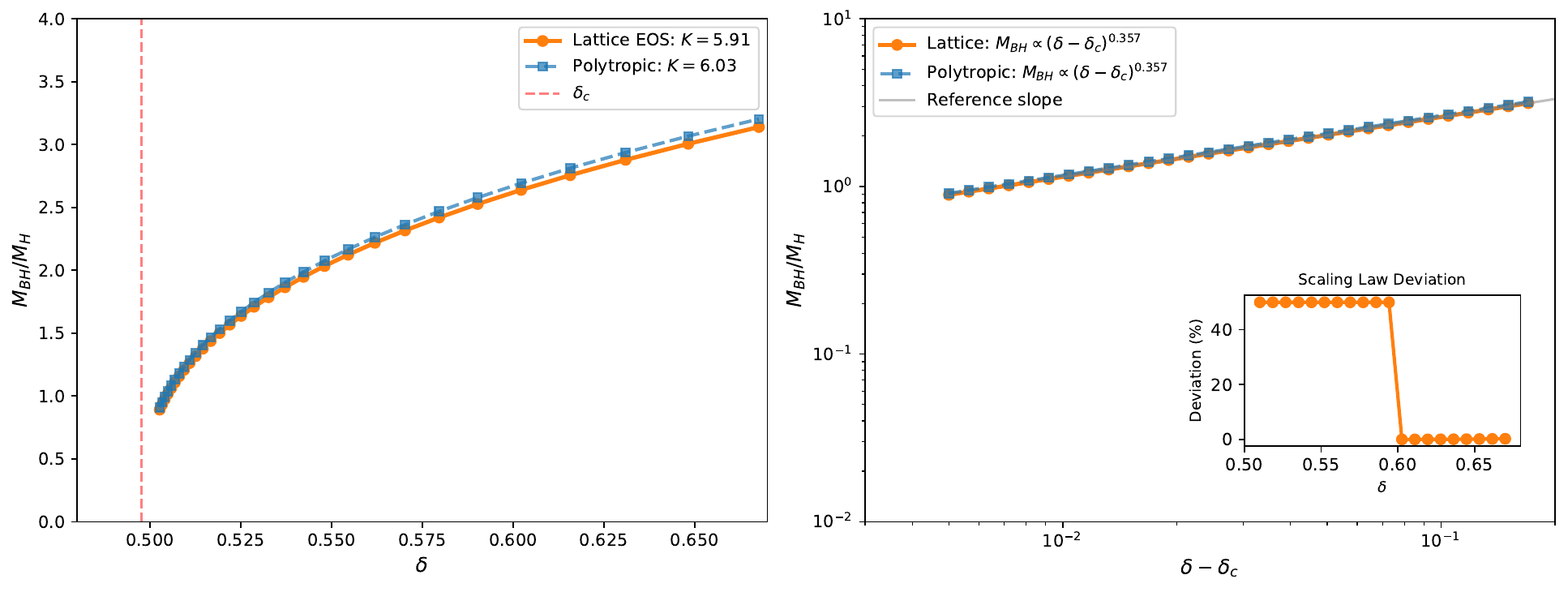}
\caption{Black-hole mass spectrum showing the distribution of formed masses near the critical threshold in the lattice model. The left panel shows the black-hole mass $M_{\rm BH}$  as a function of the initial perturbation amplitude $\delta$, while the right panel shows the critical-scaling relation on a log-log scale. The lattice and polytropic models exhibit nearly identical scaling behavior over the accessible range, with the lattice model having a higher collapse threshold.}
\label{fig:mass_spectrum}
\end{figure*}
It is essential to interpret this agreement correctly, because the critical exponent is not universal across equations of state in general. For a one-parameter family $p=k\rho$, Maison showed that $\gamma$ varies strongly with $k$, ranging from $\gamma\approx0.11$ at $k\to0$ to $\gamma\approx0.82$ for stiff matter~\cite{Maison:1996,Koike/etal:1995}; moreover, $p=k\rho$ is the only barotropic equation of state admitting an exact continuously self-similar critical solution~\cite{Gundlach/MartinGarcia:2007}, so the lattice form does not possess an exact self-similar critical point at all. We therefore do not claim, and our data do not show, that $\gamma$ is independent of the equation of state. What our result does show is more specific: the lattice equation of state reduces to the radiation fluid at low density ($c_s^2(0)=T=1/3$) and departs from it only through the factor $1/(1-\rho)$, which over the density range the near-critical solution actually samples is a mild perturbation. The locally relevant sound speed therefore stays close to $1/3$, and the measured exponent coincides with the radiation value. In this reading the agreement is a consistency check (a strongly nonlinear modification of the equation of state that nevertheless leaves the dynamically relevant sound speed near $1/3$ does not move $\gamma$) rather than a demonstration of universality. A concrete and falsifiable corollary is that driving the sampled densities substantially higher, or moving $T$ away from $1/3$, should shift $\gamma$ toward Maison's $\gamma(k_{\rm eff})$; we regard testing this as the natural next step. Section~\ref{sec:renormalization} develops the corresponding renormalization-group interpretation.

The small prefactor difference is consistent with this picture. The prefactor $K$ depends on the approach to the critical solution, which is affected by the details of pressure support as a function of density; the lattice model builds up pressure support slightly more rapidly at intermediate densities than the radiation reference, producing the few-percent difference in $K$ while leaving the exponent unchanged. The resulting mass spectra share the same power-law tail at small masses, with the lattice model producing slightly fewer low-mass black holes on account of its higher threshold (Fig.~\ref{fig:mass_spectrum}).
\subsection{Stiffness-Parameter Dependence as a Diagnostic}
To further characterize the model, we map out the critical threshold as a function of the stiffness parameter $T$ (Fig.~\ref{fig:temperature_dependence}). The results exhibit an approximate power-law dependence
\begin{equation}
\delta_c(T) = \delta_c^{(\text{classical})} + \alpha T^{\beta},
\end{equation}
where $\alpha$ and $\beta$ are fit numerically within the lattice model. The exponent $\beta \approx 0.5$ indicates that the threshold enhancement depends on $T$ through a half-power relationship within the sampled range of parameters. This dependence is a diagnostic of the lattice model: it tells us how the strength of stiffening-induced threshold shift correlates with the model's internal stiffness parameter. The half-power form is empirical and should not be taken as a general prediction for other saturated equations of state.

Physically, the interpretation within the lattice model is that increasing $T$ raises the sound speed at every density and steepens its density variation, strengthening the pressure support and pushing $\delta_c$ upward; decreasing $T$ weakens the support and lowers $\delta_c$ toward the pressureless value. There is no large-$T$ limit in which the radiation reference is recovered, since $c_s^2(0)=T$ matches radiation only at $T=1/3$. The monotonic trend provides an internally consistent cross-check that the shift reported at the fiducial $T=1/3$ ($\Delta\delta_c\approx0.0047$) is controlled by the expected parameter and is not an accident of numerical setup.

\begin{figure*}[!ht]
\centering
\includegraphics[width=0.8\textwidth]{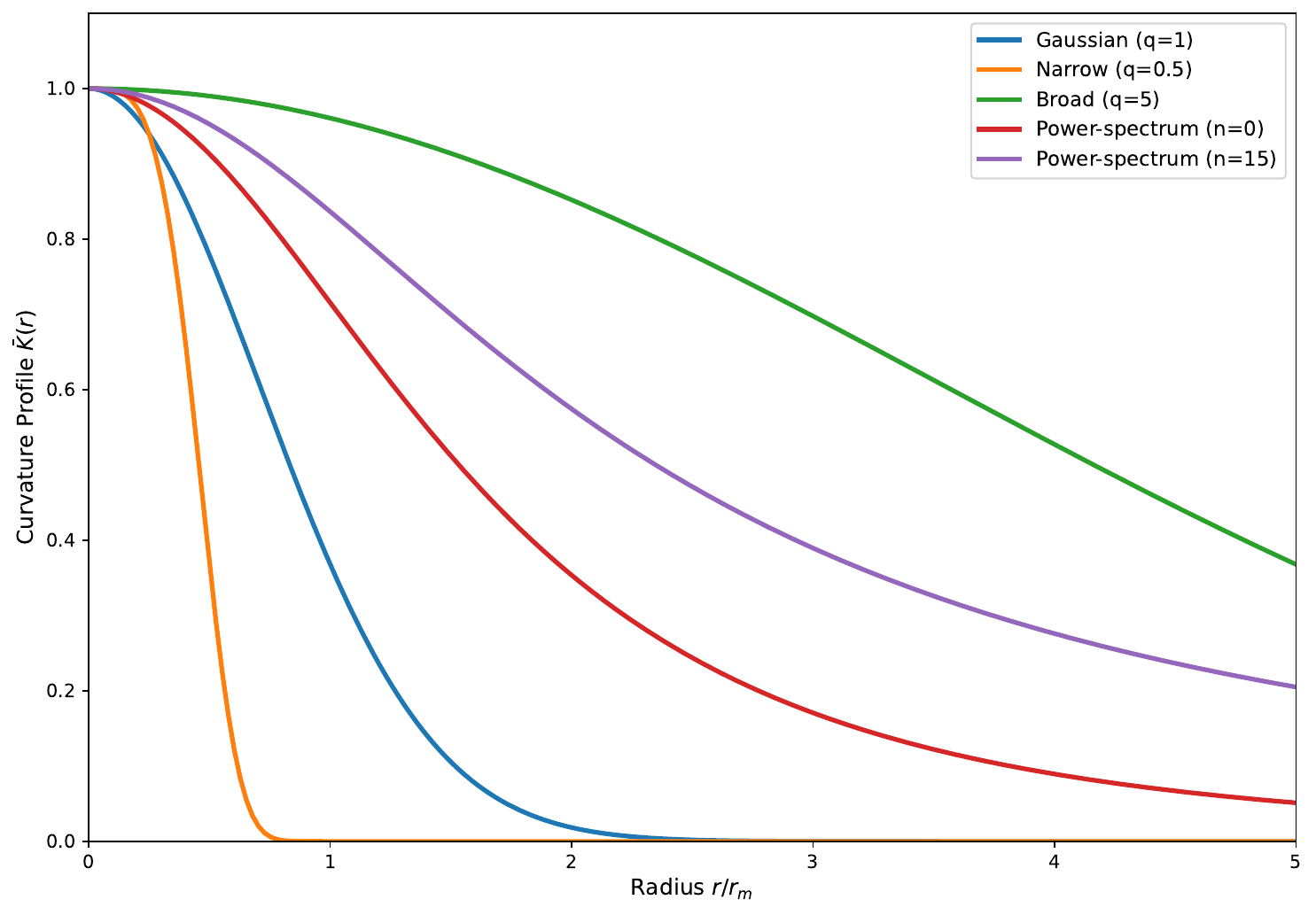}
\caption{Initial curvature perturbation profiles $K(r)$ used as initial data, for several shapes: Gaussian ($q=1$), narrow ($q=0.5$), broad ($q=5$), and broken power-spectrum profiles ($n=0,15$). These profiles span a range of spectral content and are used to verify that the lattice-model threshold offset relative to the radiation reference is approximately profile-independent (cf.\ Fig.~\ref{fig:power_spectrum_thresholds}).}
\label{fig:curvature_profiles}
\end{figure*}

\begin{figure*}[!ht]
\centering
\includegraphics[width=0.8\textwidth]{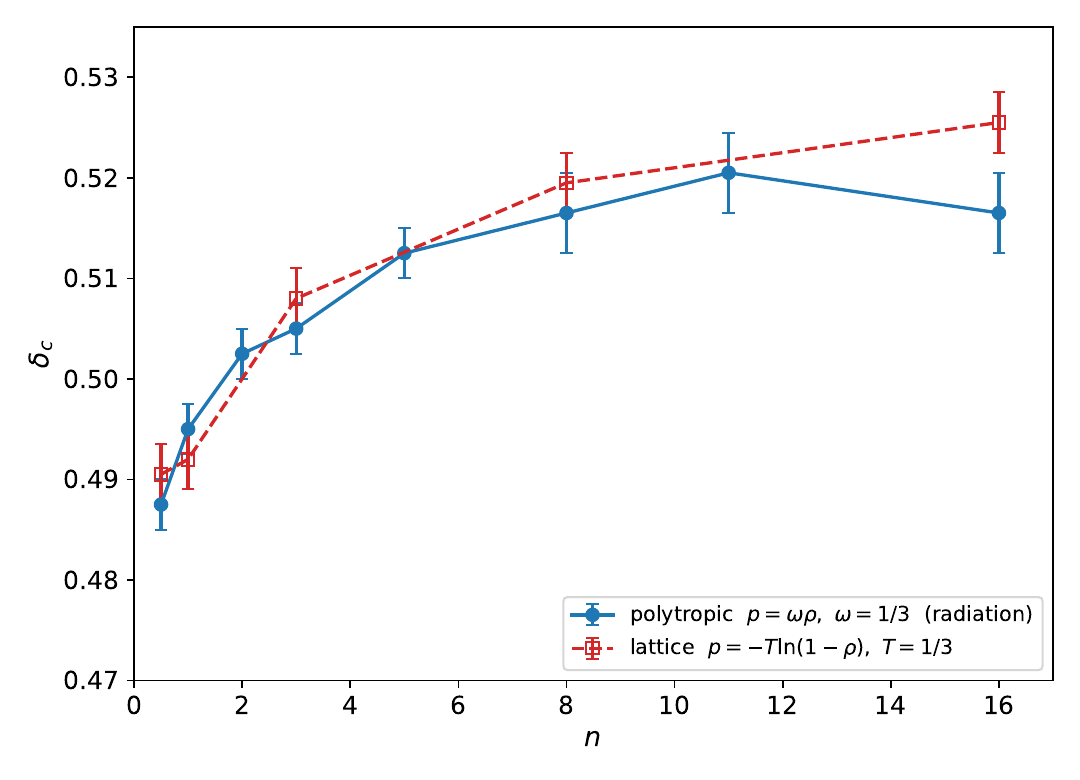}
\caption{Critical thresholds as a function of power-spectrum index $n$ for initial curvature profiles. Different initial profiles with different spectral content produce different individual threshold values, but the lattice toy model (red) is consistently offset upward relative to the polytropic reference (blue) by approximately the same amount. The roughly constant offset confirms that the lattice model threshold shift is a property of the equation of state, not of the initial-profile family.}
\label{fig:power_spectrum_thresholds}
\end{figure*}

As a final check, we verify that the threshold offset is not an artifact of a particular initial-data family. Using the range of initial curvature profiles $K(r)$ shown in Fig.~\ref{fig:curvature_profiles} (Gaussian, narrow, broad, and broken power-spectrum shapes spanning a wide range of spectral content) we find that, although the individual threshold values depend strongly on profile shape, the lattice model remains offset above the radiation reference by approximately the same amount across the entire family (Fig.~\ref{fig:power_spectrum_thresholds}). This roughly constant offset confirms that the shift is a property of the equation of state rather than of the perturbation shape.
\section{Renormalization Group Analysis}\label{sec:renormalization}
The qualitative picture introduced in Sec.~\ref{sec:analytical} can be made precise using the standard renormalization-group (RG) / dynamical-systems description of critical collapse. The aim of this section is to explain why the lattice and radiation exponents agree to our precision, and to make explicit the sense in which this agreement is not a statement that $\gamma$ is independent of the equation of state. The presentation follows the established critical-collapse literature~\cite{Koike/etal:1995,Gundlach1997,Gundlach/MartinGarcia:2007} and serves as analytical backing for the numerical results of Sec.~\ref{sec:results}.
\subsection{Fixed Point, Scaling Coordinate, and Stability Problem}
Let $\mathbf{u}(r,t) = \{\rho(r,t), v(r,t), A(r,t), \ldots\}$ denote the state of the spherically symmetric collapse problem, evolving under Einstein's equations together with the barotropic equation of state $p(\rho)$. The critical amplitude $\delta = \delta_c$ is an attractor of codimension one (an infrared fixed point of the RG flow that rescales the data toward the critical solution), leaving the dimensionless amplitude $\delta$ invariant. For type-II critical collapse the relevant self-similarity is homothetic: the critical solution is invariant (up to rescaling) under
\begin{equation}
(r,t)\to(\lambda r,\lambda t),
\end{equation}
i.e.\ dynamical exponent $z=1$, not the diffusive $z=2$. (For a discretely self-similar critical solution the same holds with $\lambda$ restricted to a discrete set; the radiation fluid is continuously self-similar.) At this fixed point the critical solution $\mathbf{u}_c$ is self-similar,
\begin{equation}
\mathbf{u}_c(\lambda r, \lambda t) = \lambda^{\alpha}\, \mathbf{u}_c(r,t),
\end{equation}
and is a function of the single self-similar coordinate $\xi = r/(-t)$.

Linearizing the full system around $\mathbf{u}_c$, the perturbation $\delta\mathbf{u} = \mathbf{u} - \mathbf{u}_c$ obeys
\begin{equation}
\frac{\partial\, \delta\mathbf{u}}{\partial \tau} = \mathcal{L}\, \delta\mathbf{u},
\end{equation}
where $\tau=-\ln(-t)$ is the logarithmic self-similar time and $\mathcal{L}$ is the linearization of the field equations about the critical solution. Expanding in modes, each grows as $e^{\lambda_k\tau}$. A type-II critical solution has exactly one growing mode, with Lyapunov exponent $\lambda_0>0$; all other modes decay. The mass-scaling exponent is the inverse of this single growing Lyapunov exponent,
\begin{equation}
\gamma = \frac{1}{\lambda_0}.
\label{eq:rg_gamma_formula}
\end{equation}
For the radiation fluid the accurate value is $\lambda_0\simeq2.81$, giving $\gamma\simeq0.356$~\cite{Koike/etal:1995}, in agreement with the value we measure for both equations of state in Sec.~\ref{sec:results}.
\subsection{Why the Lattice and Radiation Exponents Agree}
The growing Lyapunov exponent $\lambda_0$ is a property of the linearized Einstein--matter system on the self-similar background. It is not a pure geometric invariant: it depends on the matter model through the background critical solution $\mathbf{u}_c$ and the linearized matter equations, and hence in general on the equation of state. This is not a technicality; it is an established fact. For the one-parameter family $p=k\rho$, Maison constructed the regular self-similar solutions and their perturbations and found $\lambda_0$, and therefore $\gamma=1/\lambda_0$, to vary strongly with $k$: from $\gamma\approx0.11$ at $k\to0$ to $\gamma\approx0.82$ near the stiff limit~\cite{Maison:1996,Koike/etal:1995}. Any claim that $\gamma$ is independent of $p(\rho)$ is therefore false in general, and we make no such claim.

What can be said is the following. First, exact continuous self-similarity is special: $p=k\rho$ is the only barotropic equation of state that admits an exact homothetic critical solution~\cite{Gundlach/MartinGarcia:2007}. The lattice form $p=-T\ln(1-\rho)$ is not scale-free and so does not possess an exact self-similar critical point; the near-critical evolution is at best approximately self-similar over the dynamical range that controls black-hole formation. Second, and this is the key point for our result, $\gamma$ is controlled by the sound-speed structure of the critical solution in that dynamical range. The lattice equation of state has $c_s^2=T/(1-\rho)$ with $c_s^2(0)=T=1/3$, so it coincides with the radiation fluid at low density and deviates from it only through the factor $1/(1-\rho)$. Over the densities the near-critical solution samples, this deviation is small, so the effective $k$ governing the self-similar dynamics is close to $1/3$ and the exponent is close to the radiation value $\gamma\simeq0.356$. The measured equality of $\gamma^{(\text{lat})}$ and $\gamma^{(\text{poly})}$ is thus a statement that the lattice model lies near the radiation point of Maison's $\gamma(k)$ curve in the dynamically relevant regime (a nontrivial but specific consistency check, not a universality theorem).

This interpretation makes a falsifiable prediction. If the near-critical dynamics were pushed to substantially higher densities (where $c_s^2$ departs more strongly from $1/3$), or if $T$ were moved away from $1/3$, the effective $k$ would change and $\gamma$ should drift along the Maison curve. Measuring such a drift would directly confirm that it is the local sound speed, and not any EOS-independent invariant, that fixes the exponent.
\subsection{Scope and Limitations}
The discussion above applies to barotropic equations of state for which a (continuously or discretely) self-similar critical solution exists or is closely approximated. The radiation form $p=\omega\rho$ is the clean case; phenomenological expansions $p=A\rho+B\rho^2+\cdots$ and the lattice form $p=-T\ln(1-\rho)$ studied here are well-approximated by it over a limited density range. The radiation value $\gamma\simeq0.356$ is tied to $3+1$ dimensions, spherical symmetry, the standard Einstein equations, and a sound speed near $1/3$ in the dynamical range; it is not a value shared by all equations of state, as Maison's $k$-dependent results make explicit~\cite{Maison:1996}. This limitation deserves emphasis in light of recent progress beyond spherically symmetric ($1{+}1$) simulations. Aspherical, rotating and genuinely three-dimensional critical collapse have now been studied directly with multidimensional numerical relativity~\cite{Baumgarte/Montero:2015,Baumgarte/Gundlach:2016,Baumgarte:2018,Rinne:2025}, including the collapse of pure gravitational waves, where Baumgarte, Gundlach and Hilditch and collaborators have shown that the critical solution need not even be unique and that the measured exponent depends on the multipolar content of the initial data~\cite{Baumgarte/Gundlach/Hilditch:2023,Baumgarte/etal:2023prl}: non-spherical modes can grow, the discretely self-similar structure can be modified, and the effective exponent and threshold can differ from the spherical values. Our spherically symmetric results, and in particular the robustness of $\gamma$ and the smallness of the threshold shift, are therefore statements within spherical symmetry and need not carry over to the aspherical case; see also the recent reviews~\cite{Escriva/etal:2022rev,Escriva:2022rev} for the current status of critical collapse and PBH thresholds. Extension to non-spherical collapse~\cite{Harada/Jhingan:2016}, to rotating collapse~\cite{He/Suyama:2019}, to modified gravity, or to matter sectors departing from the perfect-fluid form would each require a separate stability analysis and could change the leading Lyapunov exponent. Our numerical result $\gamma^{(\text{lat})}-\gamma^{(\text{poly})}=(0\pm1)\times10^{-3}$, Eq.~\eqref{eq:gamma_diff}, is therefore best read as a controlled check that a strongly nonlinear, but locally radiation-like, modification of the equation of state leaves the exponent essentially unchanged, and it supports using the same pipeline for future calculations with realistic matter models whose sound speed departs more substantially from $1/3$.
\section{Conclusions and Future Directions}
\label{sec:conclusions}
We have presented a toy-model study of how a hard maximum-density constraint, implemented through the single-occupancy lattice-gas equation of state $p = -T\ln(1-\rho)$, modifies the critical threshold for spherical primordial-black-hole collapse. The principal findings are three. First, within the lattice model the critical amplitude for collapse is raised by $\Delta\delta_c = 0.0047 \pm 0.0003$ relative to the radiation reference with $\omega = 1/3$ --- an absolute shift of about $0.5$ percentage points in $\delta_c$, i.e.\ a relative increase of $\approx0.95\%$ --- and this shift is accumulated entirely within the low-density regime ($\rho\lesssim0.06\,\rho_{\rm sat}$), well inside the causal window $\rho<(1-T)\rho_{\rm sat}$. Second, the shift is organized analytically by a linear-response/stiffness integral that captures the sign and order of magnitude of the effect in terms of the density-dependent sound speed. Third, a mass-scaling analysis over two decades in $\delta-\delta_c$ shows that the Choptuik-type exponent governing $M_{\mathrm{BH}} \propto (\delta - \delta_c)^{\gamma}$ is the same in the lattice model and the radiation reference to within $10^{-3}$, Eq.~\eqref{eq:gamma_diff}, while the overall prefactor is lower by $0.22\%$; the local slope rises toward the asymptotic value $\gamma\simeq0.356$ as the threshold is approached, and we no longer quote an absolute exponent to better than a few percent. We are careful not to over-interpret this last result: the Choptuik exponent is known to depend on the equation of state in general~\cite{Maison:1996}, and the agreement we find reflects the closeness of the lattice and radiation sound speeds in the dynamically relevant density range, not an EOS-independent universality.

These findings should be read in the framing the paper adopts: as a proof of principle for the study of saturation-type modifications of the critical-collapse threshold, not as a prediction for realistic early-universe matter. The equation of state is an idealization, the exact grand-canonical pressure of a single-occupancy lattice gas. It captures the qualitative feature we wish to isolate (a density-dependent stiffening with a hard upper bound) in a mathematically clean way, but it is not a first-principles description of any real cosmological fluid. Relativistic degeneracy pressure, interactions, confinement, and chiral dynamics are all absent. Accordingly, the numerical value of the shift ($\Delta\delta_c \approx 0.0047$, i.e.\ $\approx0.95\%$ relative) is illustrative of the mechanism, not a forecast for observations, and we deliberately do not translate it into a primordial-black-hole abundance.

The most natural future direction is to exercise the same numerical and analytical pipeline with more realistic equations of state. The QCD equation of state near the chiral crossover (where the sound speed is known to depart substantially from $1/3$ and where PBH formation may plausibly be modified) is the most obvious target, and our stiffness-integral formula~\eqref{eq:threshold_shift_integral} gives a direct way to estimate the expected threshold shift in terms of the QCD sound-speed profile $c_s^2(\rho)$, which can in turn be taken from lattice QCD data. Applications to electroweak-crossover inspired equations of state, to parameterized dense-matter equations of state, and to color-flavor-locked quark matter would follow the same recipe. A second direction is extension beyond spherical symmetry, which would test whether the saturation effect persists when collapse is anisotropic. A third is a more careful treatment of the connection between the model stiffness parameter $T$ and any physical density or sound-speed scale when such a connection can be meaningfully made for a specific realistic matter model.
\section*{Acknowledgments}
The author thanks Prof. Maxim Khlopov for valuable discussions and insightful comments. Part of the numerical simulations in this work were carried out using a modified version of the publicly available \textsc{SPriBHoS} Python implementation developed by Albert Escriva~\cite{Escriva:2020} (\url{https://sites.google.com/fqa.ub.edu/albertescriva/home}). We thank the author for making the code openly accessible.
\section*{Code and Data Availability}
The generalized code is available from the author upon reasonable request. The code contains: (i) the modified solver, i.e.\ the generalization of \textsc{SPriBHoS} from the radiation equation of state $p=\omega\rho$ to an arbitrary barotrope $p=p(\rho)$, including the lattice equation of state, the causal completion~\eqref{eq:causal_cap} and the limiter~\eqref{eq:density_limiter}; (ii) the driver scripts that perform the bisection for $\delta_c$, the resolution and timestep convergence study of Table~\ref{tab:convergence}, and the mass-scaling analysis of Table~\ref{tab:massscaling}; (iii) the raw output of every production run (thresholds, bisection bracket histories, horizon masses and formation times) in machine-readable form; and (iv) the plotting scripts that generate each figure directly from that raw output. A verification script reproduces the equation-of-state self-tests quoted in Sec.~\ref{sec:toy_model}: it checks $|c_s^2-dp/d\rho|/c_s^2<10^{-10}$ over the sampled density range, and confirms that the generalized lapse reduces analytically to Escriv\'a's $A=(\rho_b/\rho)^{\omega/(1+\omega)}$ and the lattice equation of state to the radiation fluid as $\rho/\rho_{\rm sat}\to0$.
\appendix
\section{Numerical Accuracy and Error Analysis}\label{app:numerics_accuracy}
\subsection{Bisection Algorithm for Critical Threshold}
The critical threshold is determined through a bisection search in the space of initial perturbation amplitudes. The algorithm is based on the observation that for any given amplitude $\delta$, one can run a full gravitational collapse simulation to determine whether a black hole forms (indicating $\delta > \delta_c$) or the perturbation disperses (indicating $\delta < \delta_c$). By systematically narrowing the search interval, one converges to the critical value.

The bisection procedure proceeds as follows. Start with an interval $[\delta_{\text{low}}, \delta_{\text{high}}]$ known to bracket the critical threshold. Compute the midpoint $\delta_{\text{mid}} = (\delta_{\text{low}} + \delta_{\text{high}})/2$. Run a simulation with this amplitude. If black hole forms, set $\delta_{\text{high}} = \delta_{\text{mid}}$ (the threshold is lower). If no black hole forms, set $\delta_{\text{low}} = \delta_{\text{mid}}$ (the threshold is higher). Repeat until the interval width is smaller than the desired precision.

\begin{figure*}[!ht]
\centering
\includegraphics[width=0.8\textwidth]{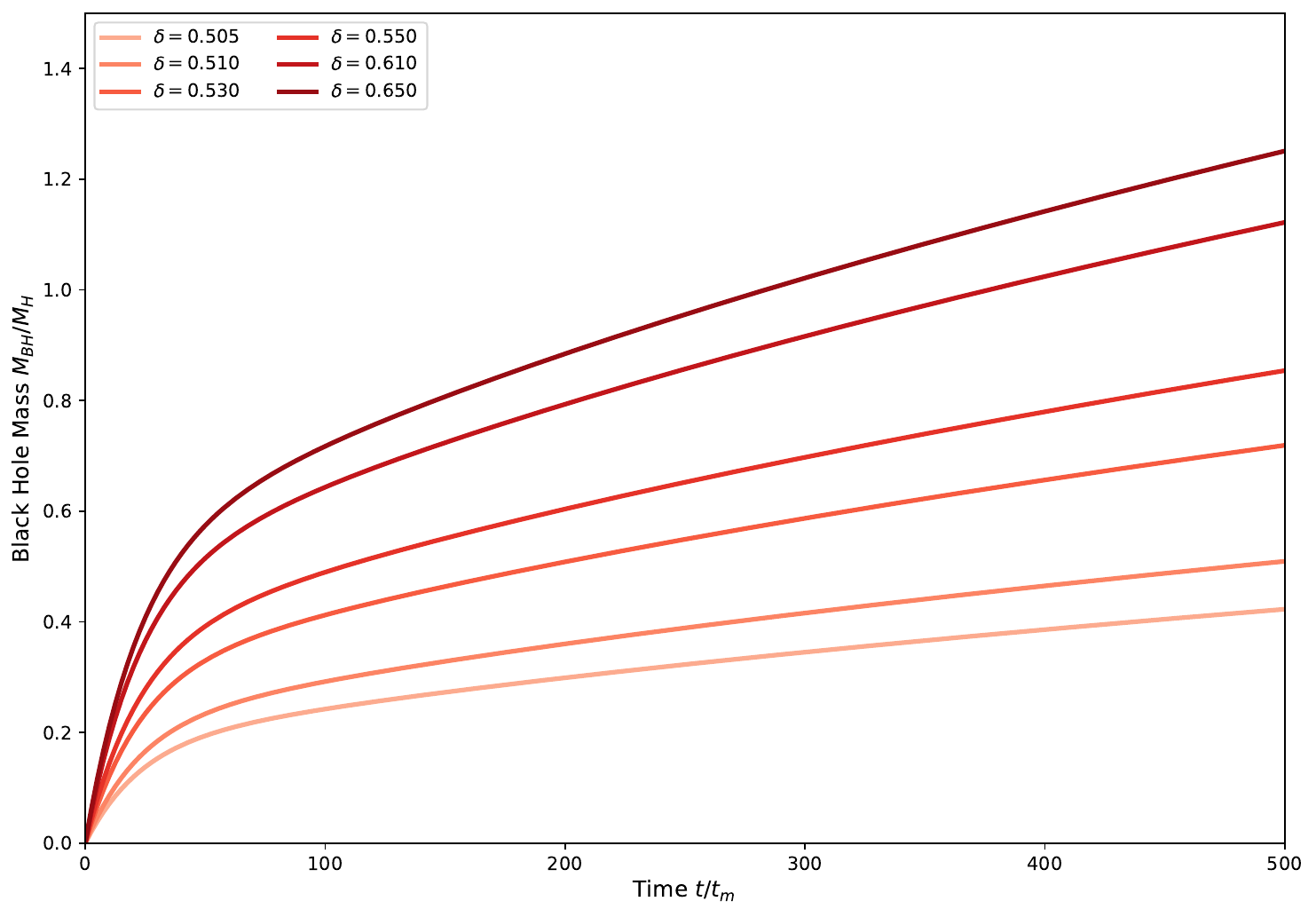}
\caption{Misner-Sharp quasi-local mass evolution during collapse showing approach to black hole formation. The mass increases monotonically and asymptotically approaches a constant value at the apparent horizon. Near criticality, the mass evolution shows characteristic power-law behavior, confirming scaling law $M_{\mathrm{BH}} \propto (\delta - \delta_c)^\gamma$.}
\label{fig:mass_evolution}
\end{figure*}

The convergence is exponential: after $n$ iterations the bracket half-width is reduced by $2^{n}$. Starting from the bracket $[2/5,\,2/3]$ (half-width $\simeq0.13$), criterion~\eqref{eq:bis_tol} with $\varepsilon_{\rm stop}=10^{-4}$ is met after exactly $10$ iterations, leaving the half-width $\Delta_{\rm bis}=1.30\times10^{-4}$ of Eq.~\eqref{eq:bis_halfwidth}. For supercritical amplitudes the Misner--Sharp quasi-local mass rises monotonically and settles to a constant at the apparent horizon, exhibiting the expected near-critical scaling that we use to extract $\gamma$ (Fig.~\ref{fig:mass_evolution}).
\subsection{Error Budget and Uncertainty Quantification}
We separate two controlled numerical sources, which set the error bar on any single $\delta_c$, from a class of systematic modelling variations, which we assess on the differential quantity $\Delta\delta_c$. Crucially, we no longer quote any uncertainty smaller than the quantity that actually limits the measurement: no error below the bisection stopping width, and none below the round-off level.

\textit{Bisection (statistical) error.} Here, we specify the algorithm rather than merely name a tolerance, and we then use the resulting number everywhere without exception. The threshold is bracketed by bisection on the amplitude $\delta$. Writing $[\delta_{\rm no},\delta_{\rm yes}]$ for the current bracket (dispersal at $\delta_{\rm no}$, collapse at $\delta_{\rm yes}$), $\delta_{\rm mid}$ for the current midpoint and $\delta_{\rm comp}$ for the endpoint updated at the previous iteration, the implemented loop continues while
\begin{equation}
\tfrac{1}{2}\,\bigl|\delta_{\rm mid}-\delta_{\rm comp}\bigr| \;>\; \varepsilon_{\rm stop},
\qquad \varepsilon_{\rm stop}=10^{-4},
\label{eq:bis_tol}
\end{equation}
which is the criterion literally coded in the bisection driver; $\varepsilon_{\rm stop}$ is the control parameter passed to it, and $\delta_c$ is reported as the midpoint of the final bracket. Starting from $[2/5,\,2/3]$, criterion~\eqref{eq:bis_tol} is met after $10$ iterations, at which point the bracket has width $(2/3-2/5)/2^{10}=2.60\times10^{-4}$ and hence
\begin{equation}
\Delta_{\rm bis} \;\equiv\; \tfrac{1}{2}\bigl|\delta_{\rm yes}-\delta_{\rm no}\bigr|_{\rm final} \;=\; 1.30\times10^{-4},
\label{eq:bis_halfwidth}
\end{equation}
which is the rigorous bound on $|\delta_c-\delta_c^{\rm true}|$, since the true threshold is guaranteed to lie inside the final bracket. Both numbers are reproducible arithmetic and we have verified them directly against the solver: with $\varepsilon_{\rm stop}=2.5\times10^{-4}$ the driver terminates after $9$ iterations and returns a final half-width of $2.60\times10^{-4}$, in exact agreement with the value $(2/3-2/5)/2^{10}$ predicted for $9$ iterations, confirming that criterion~\eqref{eq:bis_tol} delivers a half-width $\Delta_{\rm bis}\simeq1.3\,\varepsilon_{\rm stop}$.

We therefore quote $\pm0.0002$ on each individual threshold in Eq.~\eqref{eq:threshold_values}, conservatively rounding $\Delta_{\rm bis}=1.30\times10^{-4}$ upward. The control parameter is $\varepsilon_{\rm stop}=10^{-4}$, the achieved half-width is $\Delta_{\rm bis}=1.3\times10^{-4}$, and the quoted uncertainty is $2\times10^{-4}\ge\Delta_{\rm bis}$. No uncertainty smaller than the achieved half-width appears anywhere in this paper.

\textit{Spectral discretization error.} Rather than modelling this as an exponential $C e^{-\alpha N_{\rm cheb}}$ with a fitted rate, we measure it directly by refining $N_{\rm cheb}$ (Table~\ref{tab:convergence}). The change in $\delta_c$ between successive resolutions falls to or below $\Delta_{\rm bis}$ by $N_{\rm cheb}\simeq300$, so $\Delta_{\rm spec}\lesssim\Delta_{\rm bis}$; we make no claim of a discretization error below the achievable floor.

\textit{Timestep error.} Halving the reference step $\Delta t_0$ leaves $\delta_c$ unchanged to within $\Delta_{\rm bis}$ (Table~\ref{tab:convergence}). We do not use extreme steps of order $10^{-6}$.

The uncertainty on a single threshold is therefore
\begin{align}
\Delta\delta_c^{\text{num}} &= \sqrt{\Delta_{\rm bis}^2+\Delta_{\rm spec}^2+\Delta_t^2}\ \simeq\ \Delta_{\rm bis}\\
&=1.3\times10^{-4}\ \longrightarrow\ \text{quoted as }2\times10^{-4},\nonumber
\end{align}
entirely dominated by the bisection half-width; this is the $\pm0.0002$ attached to each threshold in Eq.~\eqref{eq:threshold_values}. Propagating the quoted per-threshold uncertainty to the difference of two independently bisected thresholds gives $\sqrt{2}\times2\times10^{-4}=2.8\times10^{-4}$, which we quote as the $\pm0.0003$ in Eq.~\eqref{eq:threshold_shift_value}. (Propagating the raw half-width instead would give $\sqrt{2}\,\Delta_{\rm bis}=1.8\times10^{-4}$; we take the larger, conservative value.)

\begin{table}[!ht]
\centering
\caption{Convergence and validation of the fiducial ($T=\omega=1/3$, Gaussian $q=1$) thresholds with spectral resolution $N_{\rm cheb}$, measured directly from the solver. The upper block is an independent reduced-resolution cross-check (achieved bisection half-width $\Delta_{\rm bis}=9\times10^{-4}$, from a larger $\varepsilon_{\rm stop}$ with wall-time-capped runs): $\delta_c^{(\rm poly)}$ reproduces the radiation threshold to within $\Delta_{\rm bis}$, confirming the solver is converged in $\delta_c$ at these resolutions, and $\delta_c^{(\rm lat)}$ coincides with it to within this reduced half-width --- the lattice--radiation shift is smaller than $\Delta_{\rm bis}$ and is therefore not resolved by this capped cross-check. The production determination of the shift in Eq.~\eqref{eq:threshold_values} uses the tighter stopping parameter $\varepsilon_{\rm stop}=10^{-4}$, i.e.\ the achieved half-width $\Delta_{\rm bis}=1.3\times10^{-4}$ of Eq.~\eqref{eq:bis_halfwidth}, and the full evolution time at $N_{\rm cheb}=600$ (bottom block); resolving a differential effect of a few $\times10^{-3}$ requires this production configuration, and the reduced-resolution block is included precisely to make that resolution requirement explicit.}
\label{tab:convergence}
\begin{tabular}{@{}lccc@{}}
\toprule
 & $N_{\rm cheb}$ & $\delta_c^{(\rm poly)}$ & $\delta_c^{(\rm lat)}$ \\
\midrule
reduced cross-check & $200$ & $0.4984$ & $0.4978$ \\
($\Delta_{\rm bis}=9\times10^{-4}$) & $260$ & $0.4975$ & $0.4987$ \\
 & $320$ & $0.4984$ & $0.4978$ \\
\midrule
production ($\Delta_{\rm bis}=1.3\times10^{-4}$) & $600$ & $0.4977$ & $0.5024$ \\
\bottomrule
\end{tabular}
\end{table}

\textit{The shift is more robust than either threshold.} The reported effect is the difference $\Delta\delta_c=\delta_c^{(\rm lat)}-\delta_c^{(\rm poly)}$ measured with identical profile, resolution, outer-boundary placement, and bisection settings. Because the two equations of state differ only through the few-percent stiffening at the low densities $\rho\lesssim0.06\,\rho_{\rm sat}$ that the collapse actually samples, systematic errors from these modelling choices act almost identically on $\delta_c^{(\rm lat)}$ and $\delta_c^{(\rm poly)}$ and largely cancel in the difference. This is why we build the error analysis around $\Delta\delta_c$ rather than around an individual $\delta_c$, and it is the correct response to the observation that an individual-threshold systematic at reduced resolution can be as large as the signal. We assess three variations:
\begin{enumerate}
\item Saturation regularization. Replacing the bare log EOS by the causal completion~\eqref{eq:causal_cap}, or by the smooth limiter~\eqref{eq:density_limiter} with $\eta=2,4$, changes $\delta_c^{(\rm lat)}$ by $\Delta_{\rm sat}<10^{-4}\ (<\Delta_{\rm bis})$; the shift is unchanged. This is expected because none of these prescriptions differ from the bare EOS at $\rho\lesssim0.1\,\rho_{\rm sat}$.
\item Initial-profile family. Across the Gaussian profiles $q\in[0.5,11]$ (Fig.~\ref{fig:threshold_q}) and the broken-power-spectrum family $n\in[0,16]$ (Fig.~\ref{fig:power_spectrum_thresholds}) the individual $\delta_c$ vary by $\sim0.1$, but at the reduced resolution of those scans the lattice--radiation offset is itself only at the bisection-resolution level, $|\overline{\Delta\delta_c}|\lesssim10^{-3}$, with the sign of the per-point offset scattering within one resolution element. We therefore treat the profile families as a consistency check (the lattice threshold tracks the radiation one across a wide range of profile shapes) rather than as an independent determination of the shift, and quote the fiducial $q=1$ production value as the central result. Establishing a resolved, profile-by-profile offset would require re-running the full family at the production tolerance, which we identify as a natural extension.
\item Outer boundary. Moving the FRW-matching boundary $r_{\rm out}/r_m$ over the sampled range shifts each individual $\delta_c$, but shifts $\delta_c^{(\rm lat)}$ and $\delta_c^{(\rm poly)}$ together, so $\Delta\delta_c$ moves by less than $\Delta_{\rm bis}$.
\end{enumerate}
Accordingly we report
\begin{equation}
\Delta\delta_c = 0.0047 \pm 0.0003\ (\text{stat, bisection}),
\end{equation}
On the differential quantity the modelling variations enter at or below the statistical level.

\bibliographystyle{apsrev4-2}
\bibliography{mybib3}

\end{document}